\documentclass[a4paper,11pt]{article}
\pdfoutput=1 

\usepackage{jcappub} 

\usepackage[T1]{fontenc}
\usepackage[utf8]{inputenc}
\usepackage{comment}
\usepackage{natbib}
\usepackage{hyperref}
\usepackage{multirow}
\usepackage{booktabs}

\DeclareUnicodeCharacter{2011}{-}
\DeclareUnicodeCharacter{2013}{-}
\DeclareUnicodeCharacter{2014}{--}
\DeclareUnicodeCharacter{2018}{`}
\DeclareUnicodeCharacter{2019}{'}
\DeclareUnicodeCharacter{201C}{``}
\DeclareUnicodeCharacter{201D}{''}

\def\d{\delta}

\def\s{\sigma}
\def\bd{b}
\def\bdd{\lambda}
\def\LF{\texttt{LEFTfield}}
\def\vk{\boldsymbol{k}}

\def\refeq#1{Eq.~(\ref{eq:#1})}
\def\refsec#1{Sec.~\ref{sec:#1}}
\def\reftable#1{Tab.~\ref{tab:#1}}
\def\reffig#1{Fig.~\ref{fig:#1}}
\def\refapp#1{Appendix~\ref{appendix:#1}}

\title{Field-level vs summaries: convergence of information in non-Gaussian density fields} 

\author[a]{Ivana Nikolac,}
\author[a]{Fabian Schmidt,}
\author[a,b,c]{Beatriz Tucci}

\affiliation[a]{Max-Planck-Institut f\"ur Astrophysik, Karl-Schwarzschild-Stra\ss e 1, 85748 Garching, Germany}
\affiliation[b]{Leinweber Institute for Theoretical Physics at Stanford, 382 Via Pueblo, Stanford, CA 94305, USA}
\affiliation[c]{Kavli Institute for Particle Astrophysics and Cosmology, 382 Via Pueblo, Stanford, CA 94305, USA}

\emailAdd{inikolac@mpa-garching.mpg.de}
\emailAdd{fabians@mpa-garching.mpg.de}
\emailAdd{beatucci@stanford.edu}

\abstract{
We elucidate the sources of information gain in weakly non-Gaussian cosmological fields at the field- vs. summary-statistic-level in a controlled setting.
Specifically, we compare field-level inference (FLI) with the standard power spectrum plus bispectrum (P${+}$B), and a family of \emph{composite-operator correlators} (OCs) built from auto- and cross-spectra of local powers of the galaxy density field. The forward model is a linear density field with a single local quadratic coupling $\bdd$ and Gaussian noise; this minimal nonlinear setup interpolates between a purely Gaussian dataset ($\bdd=0$) and a non-Gaussian one ($\bdd\sim 1$), while keeping the analytical structure tractable. FLI is performed by jointly sampling the initial conditions, bias and noise parameters via MCMC; the summary posteriors are obtained with simulation-based inference (SBI) as well as Fisher estimates. In the Gaussian limit, the P${+}$B, OCs and FLI yield equivalent constraints, in agreement with the perturbative expectation. As the nonlinear coupling $\bdd$ increases, the summary-based uncertainties on the model parameters grow faster than the FLI ones, leading to an increasing information loss for a fixed set of summaries. This loss is largely, but not completely, recovered by adding OCs corresponding to up to the 6-point function. The information loss over FLI becomes even more pronounced for lower-noise data, where summaries corresponding to up to the 6-point function still capture significantly less information than the field.
}

\begin{document}
\maketitle
\flushbottom

\section{Introduction}
\label{sec:intro}

When dealing with a nonlinear, non-Gaussian dataset such as the large-scale distribution of galaxies and matter, cosmologists are faced with the question of how best to extract information on the parameters of interest---usually, the parameters describing a given cosmological model such as $\Lambda$CDM or generalizations. For a Gaussian random field, the power spectrum is the \emph{sufficient statistic}. For the galaxy density field, the Gaussian assumption only holds on very large scales, and even then only applies to parameters that do not suffer from linear-order degeneracies. The most prominent such degeneracy is the bias-amplitude degeneracy: the amplitude of the linear power spectrum is entirely degenerate with the linear bias parameter $\bd \equiv b_1 \equiv b_\delta$. However, this degeneracy is broken beyond linear order by displacement contributions that are protected by the equivalence principle \cite{Desjacques_2018}.\footnote{In the presence of line-of-sight-dependent selection effects, there is a similar degeneracy of the linear growth rate $f$ \cite{Desjacques:2018pfv}.} This is just one motivation for the investigation of information beyond the power spectrum.

The question of how much cosmological information one can get by breaking the bias-amplitude degeneracy remains an active area of research. One approach is to use higher order $n$-point functions, starting with the bispectrum ($n=3$) \cite{Matarrese_1997,Scoccimarro_1998,Sefusatti_2006,Ivanov_2022,Ivanov_2023,Philcox_2022a,DAmico_2022,Philcox_2022b,Bakx_2025,Chudaykin_2025}. An alternative approach is to extract the information directly from the field, known as \emph{field-level inference (FLI)} or alternatively field-level Bayesian inference (FBI). Starting from a set of initial conditions, the corresponding evolved density field is computed (forward modeled), and compared with the data using a field-level, mode-by-mode likelihood. From this, the initial conditions, cosmological parameters and bias parameters are jointly sampled to obtain the joint posterior given the observed density field. 
Recently, Refs.~\cite{Nguyen_2024,beyond2pt} found that FLI, when applied to the case of breaking the bias amplitude degeneracy in the galaxy rest frame (i.e. without redshift-space distortions), yields substantial information gain over power spectrum and bispectrum. In similar settings but considering a higher-noise sample, Ref.~\cite{Akitsu_2025} found only moderate improvements, emphasizing the need for a careful assessment of the information gain of field-level inference relative to summary statistics.
On the other hand, analytical progress has been made in connecting the information obtainable at the field-level with $n$-point functions \cite{cabass2024cosmologicalinformation,schmidt2025connection}. Crucially, these papers performed perturbative calculations around a Gaussian field, and assumed the zero-noise limit and the absence of model mismatch, i.e. data generated from the same forward model and likelihood as used in the inference. We will review this analytical approach in \refsec{fli_intro}. Very recently, Ref.~\cite{pietroni/schmidt} presented a different formalism to directly write the FLI posterior in terms of $n$-point functions.

Our goal in this paper is to understand the quantitative comparison between field-level information and that in low-order $n$-point functions in more detail, and to connect this with the treatments in \cite{cabass2024cosmologicalinformation,schmidt2025connection}, by considering a toy scenario. Specifically, our forward model for both mock data and inference consists of a linear density field with a local quadratic interaction with coupling constant $\bdd$ and small additive Gaussian noise. In the limit of $\bdd\to 0$, we thus recover a purely Gaussian dataset, which allows us to strictly connect to the results in \cite{cabass2024cosmologicalinformation,schmidt2025connection}. On the other hand, a finite $\bdd$ generates all higher-order $n$-point functions in the data, allowing us to quantitatively test the impact of these higher-order correlators. Note that, for the above-mentioned second-order term responsible for breaking the bias-amplitude degeneracy, the corresponding coefficient is $b_1$, such that the case of interest corresponds to coupling constants $\bdd \leftrightarrow b_1$ of order unity.

Higher-order $n$-point correlation functions become increasingly difficult to handle, due to a rapid increase in the size of the data vector as well as the computational cost of the estimator. Thus, we follow a different method to access information in $(n\geq 3)$-point functions here: we adopt compressed statistics which we call \emph{composite-operator correlators}, designed, via the analytical treatments discussed above, to capture the information on $\bdd$. These statistics can be considered a special case of skew spectra \cite{Schmittfull_2015, Schmittfull:2020hoi, Dizgah_2020}.
We also compare the lowest-order composite-operator correlators with the combination of power spectrum and bispectrum, finding the expected agreement in this case.

The outline of the paper is as follows. In \refsec{fli_intro} we introduce the basics of field-level inference, with details about the forward model given in \refsec{LEFTfield}. In \refsec{zero_noise}, we calculate the maximum-a-posteriori solution for parameters of our model, which motivates the composite-operator correlators. In \refsec{summaries}, we discuss summary statistics for simulation-based inference, including composite-operator correlators and $n$-point functions. \refsec{methods} provides details on the generation of mock data, priors and specifics of both field-level inference and simulation-based inference implementation. We present inference results from both approaches in \refsec{results}, and discuss the comparison of field-level inference and summaries. Technical information is collected in the appendices, including details of the MAP calculation, the perturbative-regime criterion, the validation of the FLI and SBI posterior pipelines, the subtraction of the Gaussian disconnected contribution, and a complementary study of a lower-noise dataset.

\section{Field-level Inference}
\label{sec:fli_intro}
Field-level inference aims to extract the entire information in the
observed galaxy density field $\d_{g,\mathrm{obs}}$, by performing a full, optimal
Bayesian analysis without resorting to further compressions of the data, such as galaxy $n$-point functions. The goal of the field-level approach is to jointly infer the underlying cosmological parameters $\boldsymbol{\Omega}$, the data-specific parameters (the bias coefficients $\{b_O\}$ and stochastic amplitudes $\{\s_\varepsilon\}$), and the initial conditions $\d_\mathrm{in}$. To obtain the joint posterior given the observed galaxy density field, $\mathcal{P}(\d_\mathrm{in}, \boldsymbol{\Omega}, \{b_O\}, \{\s_\varepsilon\} \rvert \d_{g,\mathrm{obs}})$, we require a prior on the initial conditions, a forward model for the evolution of matter under gravity, a deterministic bias model, and a likelihood function \cite{Schmidt_2019}. Throughout this work, we focus on inferring only the bias and stochastic parameters, while keeping the cosmological parameters fixed. Correspondingly, we drop the dependence on $\boldsymbol{\Omega}$ in the following. We emphasize again that, following the discussion in \cite{cabass2024cosmologicalinformation,Nguyen_2024}, the inference of the primordial power spectrum amplitude from rest-frame tracers is also captured in this setup as long as all other cosmological parameters are fixed.

To describe the biased nature of galaxies as tracers of the underlying matter density field within the framework of the EFTofLSS, we can perturbatively expand the deterministic part of the galaxy density field $\d_g$ at a given point and fixed time or redshift as
\begin{equation}\label{eq:eft_density}
\d_{g,\mathrm{det}}(\vk)= \sum_O b_O O(\vk).
\end{equation}
Here, the bias operators $O$ correspond to a linearly independent set of local gravitational observables that obey the equivalence principle at a given order in perturbation theory, and $b_O$ are the corresponding bias coefficients. At each perturbative order, only a finite number of operators $O$ contribute. While the bias coefficients and operators generally depend on time or redshift, we omit the explicit dependence here, as our analysis is carried out at fixed redshift.

We forward model the bias fields in \refeq{eft_density}, starting from Gaussian initial conditions. These are determined by a single initial density field of the form
\begin{equation}\label{eq:initial_density}
\d_{\mathrm{in},\Lambda}(\vk)= W_\Lambda(k)\sqrt{P_\mathrm{L}(k)}\hat{s}(\vk)\,,
\end{equation}
smoothed at scale $\Lambda$ with a sharp-$k$ filter $W_\Lambda(k)$. Here, $P_\mathrm{L}$ is the linear power spectrum, and $\hat{s}(\vk)$ is a Gaussian random field of zero mean and unit variance.
The artificial cutoff $\Lambda$ distinguishes modes that are modeled explicitly ($k < \Lambda$) from those which are integrated out in the EFT framework ($k\geq \Lambda$), and whose effect is captured by the bias coefficients and stochasticity.

The bias expansion must also account for the stochastic effect of small-scale fluctuations on galaxy formation. After integrating out small-scale modes within the EFTofLSS framework, their effects manifest as scatter around the mean deterministic galaxy density field  $\d_{g,\mathrm{det}}$. The stochastic contributions are characterized by their $n$-point functions, where the leading order contribution $\varepsilon$ is a white Gaussian at leading order. In Fourier space, the noise power spectrum is defined through
\begin{equation}
\langle \varepsilon(\vk)\varepsilon(\vk')\rangle
= (2\pi)^3 \d_D(\vk+\vk')P_\varepsilon(k)\,,
\end{equation}
and locality of tracer formation implies that this power spectrum is constant to leading order,
\begin{equation}
P_\varepsilon(k) \approx P_{\varepsilon,0} + \mathcal{O}(k^2)\,.
\end{equation}
In this work we neglect the higher-derivative $\mathcal{O}(k^2)$ corrections arising from the non-local nature of tracer formation. The leading stochastic contribution enters the likelihood as a scale-independent real-space variance per grid cell, $\s_\varepsilon^2$. This parameter represents the amplitude of stochasticity in the large-scale limit and is related to the noise power spectrum through
\begin{equation}\label{eq:noise ps}
P_{\varepsilon,0}=\s_\varepsilon^2\frac{L^3}{(N_g^{\Lambda})^3}\,,
\end{equation} 
where $L$ is the size of the simulation box and $N_g^{\Lambda}$ is the grid size per dimension. In the remainder of the text we abbreviate $P_{\varepsilon,0}\equiv P_\varepsilon$.
Our choice of simple Gaussian white noise is motivated by the fact that we will focus on a low-noise mock dataset, where the details of the noise statistics are expected to have a small impact. It is possible to incorporate non-Gaussian noise in field-level inference, however this is substantially more costly at least in the general EFT formulation \cite{Rubira:2025rqo}.

Analytically integrating out $\varepsilon$, possible thanks to the Gaussian statistics, yields the field-level likelihood
\cite{Schmidt_2019, Cabass_2020, Rubira:2025rqo},
\begin{equation}\label{eq:eft likelihood}
\ln \mathcal{L}_\mathrm{EFT}(\d_{g,\mathrm{obs}}\rvert \hat{s}, \{b_O\}, \s_\varepsilon)=-\frac{1}{2}\sum_{\vk\neq\boldsymbol{0}}^{|\vk|< k_{\mathrm{max}}}\Bigg[\ln[2\pi \s_\varepsilon^2]+\frac{1}{\s_\varepsilon^2}\Big\rvert\d_{g,\mathrm{obs}}(\vk)-\d_{g,\mathrm{det}}[\hat{s}, \{b_O\}] (\vk)\Big\rvert ^2\Bigg]\,.
\end{equation}
The likelihood compares the observed field $\d_{g,\mathrm{obs}}$ and the deterministic model prediction $\d_{g,\mathrm{det}}$ mode by mode, up to the highest wavenumber $k_{\mathrm{max}}$ which is usually chosen to restrict the analysis to scales where the model remains reliable. Throughout this work, we set $k_{\mathrm{max}}=\Lambda$ for numerical efficiency. The mock dataset will be generated from the same model as used for the inference.

The final joint posterior of the initial density field and bias and noise parameters, with cosmological parameters fixed to fiducial values, is given by  
\begin{equation}\label{eq:joint_posterior}
\mathcal{P}(\hat{s}, \{b_O\}, \s_\varepsilon \mid \d_{g,\mathrm{obs}}) 
\propto 
\mathcal{L}_\mathrm{EFT}(\d_{g,\mathrm{obs}} \mid \hat{s}, \{b_O\}, \s_\varepsilon ) \,
\mathcal{P}(\hat{s}) \,
\mathcal{P}(\{b_O\}) \,
\mathcal{P}(\s_\varepsilon ),
\end{equation}
where we have dropped the normalizing evidence for simplicity, as is standard in Bayesian inference. The desired constraints on the bias and noise parameters can be obtained by marginalizing over the initial density field,
\begin{equation}\label{eq:marginalised_posterior}
\mathcal{P}(\{b_O\}, \s_\varepsilon \mid \d_{g,\mathrm{obs}}) \propto 
\mathcal{P}(\{b_O\}) \, \mathcal{P}(\s_\varepsilon) \int \mathcal{D} \hat{s}  \, \mathcal{P}(\hat{s}) \, \mathcal{L}_\mathrm{EFT}(\d_{g,\mathrm{obs}} \mid \hat{s}, \{b_O\}, \s_\varepsilon)\,.
\end{equation}

An analytical marginalization over the initial conditions is in general infeasible due to the nonlinearity of the forward model and the high dimensionality of the parameter space. Therefore, one needs to numerically sample the posterior in \refeq{marginalised_posterior}. Analytical progress is nevertheless possible under simplifying assumptions, as discussed in \refsec{zero_noise}.

Writing the Gaussian prior and likelihood explicitly, and restricting to Fourier modes $|\vk| < \Lambda$, \refeq{marginalised_posterior} becomes
\begin{align}
\mathcal{P}(\{b_O\}, \s_\varepsilon \mid \d_{g,\mathrm{obs}}) \propto & \int^{\Lambda} \mathcal{D} \d_{\rm in}^\Lambda \left[\prod_{\vk}^{\Lambda}  2\pi P_\mathrm{L}(k)\right]^{-1/2} 
\exp \left(-\frac{1}{2}\int_{\vk}^\Lambda\frac{|\d_{\rm in}^\Lambda(\vk)|^2}{P_\mathrm{L}(k)}\right) \nonumber \\
& \times \left[\prod_{\vk\neq\boldsymbol{0}}^{\Lambda}  2\pi \s_\varepsilon^2\right]^{-1/2} 
\exp \left(-\frac{1}{2}\int_{\vk\neq\boldsymbol{0}}^{\Lambda}\frac{|\d_{g,\mathrm{obs}}(\vk)-\d_{g,\mathrm{det}}[\d_{\rm in}^\Lambda](\vk)|^2}{\s_\varepsilon^2}\right) \nonumber \\
& \times \mathcal{P}_{\mathrm{prior}}(\{b_O\})\,\mathcal{P}_{\mathrm{prior}}(\s_\varepsilon)\,.
\label{eq:posterior}
\end{align}
The first line represents the marginalization over the initial density field, weighted by a Gaussian prior determined by the linear matter power spectrum $P_\mathrm{L}$, where we have switched the integration over $\hat s$ to one over $\d_{\rm in}^\Lambda$, the density field sharp-k filtered on the scale $\Lambda$. The second line corresponds to the EFT likelihood of \refeq{eft likelihood}, with the $\vk = 0$ mode excluded since it would be absorbed by the mean tracer density $\bar{n}$, which effectively acts as an additional free parameter. By removing this mode and working with the overdensity field $\d_g$, we can ignore $\bar{n}$.

\subsection{Forward model and LEFTfield} \label{sec:LEFTfield}
In this paper, we consider the simplest nonlinear model for structure formation, consisting of a term linear in the linearly evolved matter density and a second-order local interaction. For notational simplicity we write $\d \equiv \d_{\rm in}^\Lambda$ for the filtered linear density field in the following derivation. Including the Gaussian noise described by $\mathcal{L}_{\rm EFT}$, the data-generating process is given by:
\begin{equation}\label{eq:trivial_model}
\d_{g,\text{model}}(\boldsymbol{x}) = \d_{g,\mathrm{det}}[\d](\boldsymbol{x}) + \varepsilon(\boldsymbol{x})
 = \bd \d(\boldsymbol{x}) + \bdd[\d^2(\boldsymbol{x}) - \langle\d^2\rangle]  + \varepsilon(\boldsymbol{x})\,, 
\end{equation}
where the Gaussian noise field $\varepsilon$ has variance $\s_\varepsilon^2$, and the subtraction $\langle \d^2\rangle$ ensures $\langle \d_{g,\mathrm{det}}\rangle=0$.
In the large-scale structure context, one can identify $\bdd$ with the second-order local-in-density bias $b_{\delta^2} = b_2/2$. Crucially however, $b_{\delta^2}$ multiplies the evolved density field, while here we use a second-order coupling of the linear density field. We therefore choose a different symbol $\bdd$ to avoid any confusion.

This ``trivial'' forward model is implemented in  $\LF$, a Lagrangian, EFT-based forward model for galaxy clustering, described in detail in \cite{Schmidt:2020ovm}. Following \refeq{initial_density}, we sample a unit Gaussian field $\hat{s}(\vk)$, on a grid of size $N_g^\Lambda$ and scale it by $\sqrt{P_\mathrm{L}}$ to obtain the initial linear density field $\d$ at the target redshift, which we fix to $z=0$ throughout. Given the box size $L$, the grid size is chosen so that all Fourier modes up to $\Lambda$ are represented, and the cutoff is imposed by a sharp-$k$ filter. The quadratic operator $\d^2$ is then constructed from $\d$ in real space, on a suitably enlarged grid to avoid aliasing.

\subsection{Zero-noise limit}\label{sec:zero_noise}

The integral appearing in \refeq{posterior} is generally difficult to evaluate analytically, requiring numerical sampling of the initial conditions. Analytical progress is possible, however, for the case $k_\mathrm{max}=\Lambda$ in the zero-noise limit, as discussed in detail in~\cite{cabass2024cosmologicalinformation, schmidt2025connection}.

First, setting $k_\mathrm{max}=\Lambda$ ensures that the modes included in the likelihood are in one-to-one correspondence with those in the initial conditions. Second, to isolate the maximum-a-posteriori (MAP) solution, we take the zero-noise limit $P_\varepsilon \to 0$. While this assumption is not physically realistic, it allows us to gain intuitive and quantitative understanding of our problem and provide a controlled setting for the analysis of the information content in the galaxy density field. In this limit, the likelihood in the second line of \refeq{posterior} reduces to a Dirac delta functional centered on $\d_g^{\Lambda}$, and the posterior becomes
\begin{align}
\mathcal{P}[\bd,\bdd|\d_g] \propto & \int^{\Lambda} \mathcal{D} \d \left[\prod_{\vk}^{\Lambda}  2\pi P_\mathrm{L}(k)\right]^{-1/2} 
\exp \left(-\frac{1}{2}\int_{\vk}^\Lambda\frac{|\d(\vk)|^2}{P_\mathrm{L}(k)}\right) \nonumber \\
& \times \d_\mathrm{D}^{[0, \Lambda]}(\d_g-\d_{g,\mathrm{det}}[\d]) \times \mathcal{P}_{\mathrm{prior}}(\bd,\bdd)\,.
\label{eq:posterior_zero}
\end{align}

The Dirac delta functional imposes $\d_{g,\mathrm{det}}[\d]=\d_g$, which for our quadratic forward model \refeq{trivial_model} has two solutions. Dropping the tadpole subtraction $-\bdd \langle\d^2\rangle$, which is equivalent to excluding the $\vk=0$ mode, these are
\begin{align}
\d_{g,\mathrm{det}}^{-1}[\d_g, \bd, \bdd](\boldsymbol{x}) = \frac{-\bd \pm \sqrt{\bd^2+4\bdd\d_g(\boldsymbol{x})}}{2\bdd}\,.
\label{eq:two_solution}
\end{align}
Here, the data $\d_g \equiv \d_g^\Lambda$ is sharp-k filtered on the scale $\Lambda$.
We select the branch that connects smoothly to the linear inversion $\d = \d_g/\bd$ in the limit $\bdd\to 0$, yielding the perturbative expansion
\begin{align}
\d_{g,\mathrm{det}}^{-1}[\d_g, \bd, \bdd](\boldsymbol{x}) = \frac{1}{\bd}\d_g(\boldsymbol{x})-\frac{\bdd}{\bd^3}\d_g^2(\boldsymbol{x})+\mathcal{O}(\d_g^3)\,.
\label{eq:inverse_solution}
\end{align}
With this single-value inverse, the path integral over $\d$ in \refeq{posterior_zero} can be performed, giving
\begin{align}
\mathcal{P}[\bd, \bdd|\d_g] \propto & \left[\prod_{\vk}^{\Lambda}  2\pi P_\mathrm{L}(k)\right]^{-1/2} 
\exp \left(-\frac{1}{2}\int_{\vk}^\Lambda\frac{|\d(\vk)|^2}{P_\mathrm{L}(k)}\right)\Bigg\rvert_{\d=\d_{g,\mathrm{det}}^{-1}[\d_g, \bd, \bdd]} \nonumber\\
& \times \Bigg\rvert \frac{\mathcal{D}\d_{g,\mathrm{det}}}{\mathcal{D}\d}\Bigg\rvert_{\d_{g,\mathrm{det}}^{-1}[\d_g, \bd, \bdd]}^{-1} \mathcal{P}_{\mathrm{prior}}(\bd,\bdd)\,,
\label{eq:posterior_integrated}
\end{align}
where the Jacobian arises from the change of variables $\d \to \d_{g,\mathrm{det}}[\d]$ in the Dirac delta functional. Taking the logarithm yields
\begin{align}
-2\ln \mathcal{P}[\bd, \bdd|\d_g] =  & \int_{\vk}^\Lambda\frac{|\d_{g,\mathrm{det}}^{-1}[\d_g, \bd, \bdd](\vk)|^2}{P_\mathrm{L}(k)} +2\ln \Bigg\rvert \frac{\mathcal{D}\d_{g,\mathrm{det}}}{\mathcal{D}\d}\Bigg\rvert_{\d_{g,\mathrm{det}}^{-1}[\d_g, \bd, \bdd]} \nonumber \\
& +\ln \left[\prod_{\vk}^{\Lambda}  2\pi P_\mathrm{L}(k)\right]-2 \ln \mathcal{P}_{\mathrm{prior}}(\bd,\bdd) + \mathrm{const.}\,.
\label{eq:log_posterior}
\end{align}
The normalization $\ln \left[\prod_{\vk}^{\Lambda}  2\pi P_\mathrm{L}(k)\right]$ is independent of the model parameters $\{\bd, \bdd\}$ we aim to infer and can be absorbed into the constant. Moreover, assuming uniform priors implies $-2\ln \mathcal{P}_{\mathrm{prior}}(\bd,\bdd)=\mathrm{const.}$, which can likewise be dropped. Including informative priors in the following would be straightforward, but would not add any new insights.

The maximum-a-posteriori (MAP) estimate for $\bdd$ given the data $\d_g$, keeping $\bd$ fixed can be determined as (for a detailed derivation, see \refapp{MAP})
\begin{align}
\bdd & = \frac{N[\d_g]}{D[\d_g]} \quad \mathrm{with }\nonumber \\
N[\d_g] & =\frac{1}{\bd^2}\int_{\vk}^\Lambda\frac{1}{P_\mathrm{L}(k)}(\d_g^2)(-\vk)\d_g(\vk), \nonumber \\
D[\d_g] & =\frac{1}{\bd^4}\int_{\vk}^\Lambda\frac{1}{P_\mathrm{L}(k)}(\d_g^2)(-\vk)(\d_g^2)(\vk) - \frac{8}{\bd^2}\int_{\boldsymbol{x}}\d_g^2(\boldsymbol{x})\,.
\label{eq:b_2_map}
\end{align}
The numerator $N[\d_g]$ is a weighted integral over the bispectrum, while the denominator $D[\d_g]$ contains a weighted integral over the 4-point function (the first term, including both connected and disconnected parts) and a contribution from the Jacobian (the second term).

From the inverse solution \refeq{inverse_solution}, the next term in the expansion is $(2\bdd^2/\bd^5)\d_g^3$ (see \refapp{MAP}), which introduces 5-point and 6-point functions into the MAP expression for $\bdd$ in \refeq{b_2_map}. In general, the MAP expressions become increasingly complex at higher orders, and contain products of different operators, as discussed in detail in \cite{schmidt2025connection}. However, it is clear that the field-level information on $\bdd$ in the zero-noise limit is encoded in auto- and cross-correlations of local powers of the filtered data, $[\d_g^\Lambda(\boldsymbol{x})]^n$.

\section{Summary statistics}
\label{sec:summaries}

\subsection{Composite-Operator Correlators}

The standard cosmological inference procedure relies on compressing the galaxy density field into a set of lower-dimensional summary statistics that are informative about the parameters of interest. The structure of the MAP estimate for $\bdd$ in \refeq{b_2_map} includes correlations of powers of the observed galaxy density field. This motivates the introduction of \emph{composite-operator correlators} (OCs), where the summary statistics consist of a set of cross- and auto-spectra of operators $O^{(n)}[\d_g^\Lambda]$ applied to the galaxy density field
\begin{equation}
  \Big\{ \langle O^{(n)}[\d_g^\Lambda] \, O^{(m)}[\d_g^\Lambda] \rangle \Big\}_{n,m}\,,
  \label{eq:OCdef}
\end{equation}
where in this work we restrict ourselves to the local power operators $O^{(n)}[\d_g] \equiv [\d_g]^n$. In the following, we will drop the explicit filtering scale $\Lambda$, as we will exclusively work with the filtered data.

For $n=m=1$, \refeq{OCdef} reduces to the standard galaxy power spectrum. For $n=1, m=2$, the correlator corresponds to an integral over the 3-point function or bispectrum of the data, while for $n=m=2$, it is a double integral over the 4-point function (including disconnected contributions). Note that the set of correlators with $n,m \in \{1, 2\}$ corresponds exactly to the structure appearing in the field-level MAP estimate of $\bdd$ derived in \refsec{zero_noise}. In the zero-noise limit considered here, OCs should therefore contain the complete information up to the chosen order in $\d_g$, allowing the MAP estimate of $\bdd$ to be computed without evaluating the full high-dimensional field integrals. We emphasize that the set in \refeq{OCdef} is only complete for a data model with simple local couplings, as is the case for the forward model \refeq{trivial_model}. For a nontrivial forward and bias model, one needs to include additional nonlocal operators of the data \cite{Schmittfull_2015, Schmittfull:2020hoi, Dizgah_2020, schmidt2025connection}. We leave the analysis of such operators and their corresponding OCs to future work.

In practice, before measuring the cross-spectra we Wick-order each operator to remove pieces that are redundant with the power spectrum: at $n=2$ we use $\d_g^2 - \langle\d_g^2\rangle$, and at $n=3$ we use $\d_g^3 - 3\langle\d_g^2\rangle\,\d_g$. With these subtractions, the $\vk=0$ mode is excluded from every measured correlator. We keep the abbreviated notation $\langle\d_g^n\,\d_g^m\rangle$ throughout for the resulting cross-spectra. The Gaussian disconnected part of $\langle\d_g^2\,\d_g^2\rangle$ is additionally subtracted at the level of the measured spectrum, as described in \refapp{disconnected}.

Our approach is closely related to that of skew-spectra \cite{Schmittfull_2015,Dizgah_2020}, which likewise compress higher-order $n$-point functions (see \refsec{PB}) into cross-spectra of operators. While skew-spectra cross-correlate the galaxy density field with weighted quadratic fields built from $\d_g$, with kernels typically chosen to match the tree-level bispectrum, OCs in our definition include \emph{all auto- and cross-spectra of operators} up to a maximum order $n$, allowing them to systematically capture higher-order correlations. More recently, the skew-spectrum idea has been extended to composite-field galaxy correlators \cite{gao2025paritycompositefieldgalaxycorrelators}, which similarly measure cross-correlations of arbitrary nonlinear transformations of the density field. OCs can be viewed as a special case of these composite-field correlators.

\subsection{\texorpdfstring{Galaxy $n$-point functions}{Galaxy ``n''-point functions}}
\label{sec:PB}
A familiar way to construct summary statistics from the galaxy density field is through its $n$-point correlation functions.

The galaxy power spectrum $P_g$ ($n=2$), bispectrum $B_g$ ($n=3$), and trispectrum $T_g$ ($n=4$) are defined via \cite{Desjacques_2018}:
\begin{align}
\label{eq:PBT}
 \langle \d_g (\mathbf{k}) \, \d_g(\mathbf{k}')\rangle
 &= (2\pi)^3 \, \d_D(\mathbf{k}+\mathbf{k}') \, P_g(k),\\[6pt]
 \langle \d_g (\mathbf{k}_1) \, \d_g (\mathbf{k}_2)\, \d_g (\mathbf{k}_3)\rangle
 &= (2\pi)^3 \, \d_D(\mathbf{k}_1+\mathbf{k}_2+\mathbf{k}_3) \, B_g(k_1, k_2, k_3),\\[6pt]
\langle \d_g(\mathbf{k}_1)\d_g(\mathbf{k}_2)\d_g(\mathbf{k}_3)\d_g(\mathbf{k}_4)\rangle
&= (2\pi)^3 \d_D(\mathbf{k}_1{+}\mathbf{k}_2{+}\mathbf{k}_3{+}\mathbf{k}_4)\, T_g(\mathbf{k}_1,\mathbf{k}_2,\mathbf{k}_3,\mathbf{k}_4)\notag \\[6pt]
&\quad + (2\pi)^6\Big[
\d_D(\mathbf{k}_1+\mathbf{k}_2)\,\d_D(\mathbf{k}_3+\mathbf{k}_4)\,P_g(k_1)\,P_g(k_3)
\nonumber\\
&\qquad\qquad\qquad\;+
\d_D(\mathbf{k}_1+\mathbf{k}_3)\,\d_D(\mathbf{k}_2+\mathbf{k}_4)\,P_g(k_1)\,P_g(k_2)
\nonumber\\
&\qquad\qquad\qquad\;+
\d_D(\mathbf{k}_1+\mathbf{k}_4)\,\d_D(\mathbf{k}_2+\mathbf{k}_3)\,P_g(k_1)\,P_g(k_2)
\Big]\,.
\end{align}
The four-point function is included here for completeness, as it appears in the discussion of the OC correlator $\langle \d_g^2 \, \d_g^2 \rangle$. For our inference, we employ the bispectrum estimator introduced in \cite{Scoccimarro_1998}, which also yields the power spectrum as a byproduct. We do not perform an inference using the trispectrum.

\subsection{\texorpdfstring{Mapping between $n$-point functions and composite-operator correlators}{Mapping between ``n''-point functions and composite-operator correlators}}
\label{sec:OCvsPB}

Composite-operator correlators (OCs) provide a convenient way of capturing higher-order correlations of the galaxy field while avoiding the complexity of measuring $n$-point functions directly. At each order $n$, the set of operators $\{O^{(n)}[\d_g]\}$ that enters the OCs is constructed to include the operators that appear in the EFT forward model for the deterministic part of the $n$-th order galaxy density field.
It is straightforward to see that at $n=2$, the OCs $\langle \d_g \, O^{(2)}[\d_g]\rangle$ capture the configuration dependence of the tree-level bispectrum. At next order, the set $\{\langle \d_g \, O^{(3)}[\d_g]\rangle,\  \langle O^{(2)}[\d_g] \, O'^{(2)}[\d_g]\rangle\}$ captures the complete configuration dependence of the tree-level connected galaxy 4-point function (trispectrum). 

\begin{figure}[tbp]
\centering 
\includegraphics[width=.80\textwidth]{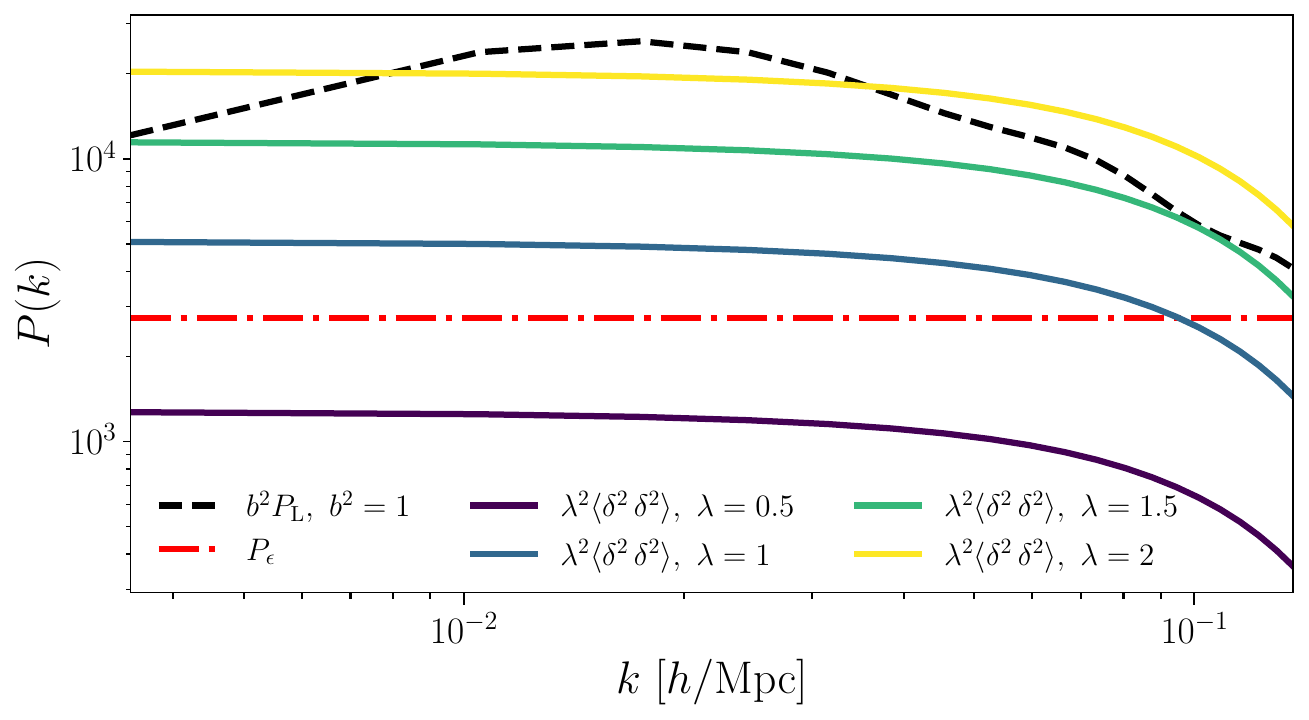}
\caption{\label{fig:components} Contributions to the galaxy power spectrum $\langle \d_g\ \d_g\rangle$ from the forward model \refeq{trivial_model} evaluated at cutoff $\Lambda = 0.14\,h/\mathrm{Mpc}$. The black dashed curve shows the linear bias term $\bd^2 P_\mathrm{L}$ for $\bd=1$, and the solid colored curves show the nonlinear contribution $\bdd^2 \langle \d^2 \,\d^2 \rangle$ for increasing values of $\bdd$, where $\d \equiv \d_{\rm in}^\Lambda$ denotes the linear density field sharp-k filtered on the scale $\Lambda$. The red dash-dotted curve indicates the stochastic noise level $P_\varepsilon$ at the fiducial value $\s_\varepsilon=0.5$. Note that the cross term $2\bd\bdd \langle \d \,\d^2 \rangle$ vanishes as $\d$ is a Gaussian field.}
\end{figure}

The relative importance of higher-order OCs depends on the strength of the nonlinearities in the galaxy field, controlled in our forward model \refeq{trivial_model} by $\bdd$. \reffig{components} illustrates this by showing the two deterministic contributions to the galaxy power spectrum $\langle \d_g\, \d_g\rangle$: the linear bias contribution $\bd^2 P_\mathrm{L}$ and the nonlinear coupling contribution $\bdd^2 \langle \d^2\, \d^2 \rangle$ (recall that $\d$ is evolved linearly), both evaluated at the cutoff $\Lambda = 0.14 \,h/\mathrm{Mpc}$, together with the stochastic noise level $P_{\varepsilon}$ from \refeq{noise ps}. The two deterministic contributions differ in both amplitude and scale dependence: the nonlinear term scales as $\bdd^2$ and becomes increasingly important toward smaller scales, while its shape reflects the convolution structure of the squared field, regulated by $\Lambda$. We use $\s_\varepsilon=0.5$ as fiducial value, such that the noise level $P_\varepsilon$ stays below the total deterministic signal across all scales and for all values of $\bdd$ considered, including $\bdd=0$ where only the linear term contributes. 

The comparison between FLI and summaries we present in \refsec{results} remains valid as a controlled test for any value of $\bdd$ in the context of the exact model \refeq{trivial_model}. However, if \refeq{trivial_model} is interpreted as the lowest-order terms of an EFT expansion rather than as an exact model, the model is only applicable if higher-order terms are consistently suppressed. In this interpretation, which is the one relevant to real-world EFT-based analyses, the regime of validity of \refeq{trivial_model} depends on the value of $\bdd$: as $\bdd$ grows, the cutoff $k_\mathrm{nl}$ at which the data become fully nonlinear decreases.
Keeping $k_\mathrm{max}\equiv\Lambda$ fixed and varying the degree of nonlinearity, we quantify this through the ratio of the leading 1-loop contribution to the tree-level power spectrum
\begin{equation}\label{eq:nl_ratio}
R(\bdd) \equiv \max_{k < \Lambda}\,\frac{P_\mathrm{1-loop}(k|\bdd)}{P_\mathrm{tree}(k)}=\max_{k < \Lambda}\,
\frac{\bdd^2\,\langle\d^2(\vk)\,\d^2(-\vk)\rangle}{\bd^2\,P_\mathrm{L}(k)}\,,
\end{equation}
where we use the maximum over $k$, rather than the value at $\Lambda$, to obtain a conservative estimate. When $R \ll 1$, the 1-loop correction is small and the model is perturbatively controlled, i.e. the effect of higher-order terms neglected in \refeq{trivial_model} should be consistently suppressed; $R = 1$ marks the breakdown of the expansion. \reffig{linear_regime} in \refapp{linear_regime} shows $R(\bdd)$ explicitly.

Computationally, evaluating all cross-spectra of the relevant operators is cheap, which makes OCs more efficient than measuring all $n$-point configurations directly. Consider a set of $N_{\rm bin}$ wavevector bins in which each of the composite-operator correlators is measured. Including all correlators up to $n,m \leq M$ leads to a total data vector of length $M(M+1) N_{\rm bin}/2$. We can therefore easily consider the full set of OCs consisting of all auto- and cross-power spectra of operators up to a chosen maximum order $n$. On the other hand, the $(M+1)$-point function of the data alone has dimensionality $\sim N_{\rm bin}{}^M$, an exponential scaling which quickly overwhelms the OC data vector.

Based on the connection between OCs and $n$-point functions, in this paper we consider the following OC combinations that capture increasing levels of statistical information from the galaxy field:
\begin{itemize}
    \item OC$\_$2nd$\_$LO: $\{\langle \d_g \,\d_g\rangle, \langle \d_g \,\d_g^2\rangle\}$ $\to$ power spectrum (P) + bispectrum (B)
    \item OC$\_$2nd$\_$full: $\{\langle \d_g \,\d_g\rangle, \langle \d_g \,\d_g^2\rangle, \langle \d_g^2 \,\d_g^2\rangle\}$ $\to$ adds part of trispectrum (T)
    \item OC$\_$3rd$\_$LO: $\{\langle \d_g \,\d_g\rangle, \langle \d_g \,\d_g^2\rangle, \langle \d_g^2 \,\d_g^2\rangle, \langle \d_g \,\d_g^3\rangle\}$ $\to$ P + B + T
    \item OC$\_$3rd$\_$full: $\{\langle \d_g \,\d_g\rangle, \langle \d_g \,\d_g^2\rangle, \langle \d_g^2 \,\d_g^2\rangle, \langle \d_g \,\d_g^3\rangle, \langle \d_g^2 \,\d_g^3\rangle, \langle \d_g^3 \,\d_g^3\rangle\}$ $\to$ adds part of 5- and 6-point functions.
\end{itemize}
While the expressions above are written as expectation values, in practice OCs are estimated as sums over Fourier modes in bins, analogous to standard power spectrum estimators.

In the linear limit $\bdd \to 0$, the galaxy field reduces to $\d_g = \bd\,\d$ and is Gaussian, so all higher-order correlators either vanish or are fully determined by the power spectrum. In this limit, OC$\_$2nd$\_$LO, the P${+}$B combination, and field-level inference should yield equivalent constraints, providing a useful consistency check that we verify in \refsec{results}.

When calculating the auto-correlation of the squared density field $\d_g^2$, we are effectively probing the four-point function, which decomposes into a connected part containing the true non-Gaussian correlations, and a disconnected part built from products of two-point functions. To isolate the connected part, which contains the physical information we want, we subtract the disconnected part when evaluating $\langle \d_g^2 \,\d_g^2\rangle$. The details of implementing this in the code are given in \refapp{disconnected}. Note that this was done to improve the efficiency of the SBI pipeline, and that the posterior should ultimately not be affected.

We compare parameter constraints from OCs to those obtained from the galaxy power-spectrum and bispectrum. We measure these $n$-point functions on the same grids as the OCs, generated with $\LF$.

\section{Methods}\label{sec:methods}

Our goal in this paper is to compare the posteriors for the model parameters $\bd$, $\bdd$, and $\s_\varepsilon$ obtained from field-level inference (FLI), composite-operator correlators, and $n$-point functions. FLI employs MCMC sampling, while for the summary statistics we use simulation-based inference (SBI). SBI does not rely on analytical predictions for the mean and covariance for the summaries, but instead these are directly derived from the same forward model used in FLI and for the mock data generation. This ensures a consistent comparison between field-level and summary-statistic-based inference. Finally, we compute Fisher forecasts for the summary statistics, where the mean and covariance of the data vector are likewise derived from realizations of the same forward model. Unlike SBI, the Fisher forecast approximates both data vector likelihood and parameter posterior as Gaussian, and generally underestimates the error bar. Still, it provides a good cross-check of the SBI result. We describe the three approaches below.

\subsection{Data sets}
\label{sec:mocks}
The data sets, or ``mocks'', used in this paper for both FLI and SBI are generated with $\LF$. All mocks are produced in a cubic box of side length $L = 2000~h^{-1}\mathrm{Mpc}$, assuming a fiducial $\Lambda$CDM cosmology with parameters $\Omega_m = 0.3$, $\Omega_\Lambda = 0.7$, $h = 0.7$, and $n_s = 0.967$. A sharp-$k$ cutoff $\Lambda = 0.14 \,h/\mathrm{Mpc}$ is applied, setting the grid size correspondingly to $N_g^\Lambda = 90$, so that the Nyquist frequency of the grid matches the cutoff for our box size.

For all mocks we fix the linear bias parameter to $\bd^{\mathrm{fid}} = 1$. Nonlinearity is introduced through the quadratic coupling parameter $\bdd$ in the data model \refeq{trivial_model}. We consider the fiducial values
\begin{equation}
\bdd^{\mathrm{fid}} \in \{0,\;0.25,\;0.5,\;0.75,\;1,\;1.25\}.
\end{equation}
We do not consider negative values of $\bdd$, since the leading contributions to the relevant summary statistics scale as even powers of $\bdd$, implying a symmetry around $\bdd=0$.

We set the stochastic amplitude to $\s^{\mathrm{fid}}_\varepsilon = 0.5$ for most mocks, corresponding to the Poisson shot noise of tracers with comoving number density $\bar{n} \approx 3.65 \times 10^{-4} (h^{-1}\mathrm{Mpc})^{-3}$. A lower noise case, $\s^{\mathrm{fid}}_\varepsilon = 0.3$, was also tested for consistency (see \refapp{lower_noise}).

\subsection{Field-level inference}
We perform field-level inference to constrain the model parameters $\bd$, $\bdd$, and the stochastic amplitude $\s_\varepsilon$ using the EFT likelihood defined in \refeq{eft likelihood}. As noted in \refsec{fli_intro}, all cosmological parameters are fixed to their fiducial $\Lambda$CDM values. The FLI parameter space is therefore $\{\hat{s}, \bd, \bdd, \s_\varepsilon\}$, where $\hat{s}$ is a three-dimensional grid of size ${(N_g^\Lambda)}^3$.

To efficiently evaluate the posteriors, we treat $\hat{s}$ and the bias and noise parameters separately. The high-dimensional field $\hat{s}$ ($N_\mathrm{dim} \sim 10^6$) is sampled with Hamiltonian Monte Carlo (HMC). HMC exploits gradient information to propose long, correlated steps, so the cost per independent sample scales as $N_\mathrm{dim}^{1/4}$, compared to $N_\mathrm{dim}$ for classical Metropolis-Hastings MCMC \cite{Brooks_2011}. 
The bias parameters and stochastic amplitude are sampled with sequential univariate slice sampling, drawing each parameter in turn from its one-dimensional conditional distribution given the current realization of $\hat{s}$. 

For each fiducial $\bdd$ value, we run five chains: one true-phase initialization (TPI) chain initialized at the ground-truth initial conditions from the mock, and four random-phase initialization (RPI) chains initialized from independent random phases. For each chain, we estimate the integrated auto-correlation time $\hat\tau_i$ of every sampled parameter $i$ and discard the first $5\hat\tau_{\rm max}$ samples as burn-in, with $\hat\tau_{\rm max}$ taken as the maximum across $\{\hat\tau_\bd, \hat\tau_\bdd, \hat\tau_{\s_\varepsilon}\}$. The post-burn-in segments of all five chains are then concatenated into a single combined chain, which we extend until it contains at least 100 effective samples of $\bdd$. Convergence is verified with the Gelman--Rubin statistic $\hat R$. Further details and per-mock diagnostics are given in \refapp{FLI_validation}.

\subsection{Simulation-based inference of summary statistics}
Simulation-based inference (SBI) estimates parameters by training on forward-model simulations, thereby learning the posterior without requiring an explicit likelihood. Let $\boldsymbol{\theta} \in \mathbb{R}^{N_\theta}$ denote the parameter vector (e.g., bias and noise parameters) and $\boldsymbol{x} \in \mathbb{R}^{D}$ the data vector (e.g., power spectrum or operator correlator bins). Using a proposal distribution $\tilde{p}(\boldsymbol{\theta})$ and a simulator for $p(\boldsymbol{x}|\boldsymbol{\theta})$, we generate $N_{\mathrm{sim}}$ joint samples $(\boldsymbol{\theta}_n, \boldsymbol{x}_n) \sim p(\boldsymbol{\theta}, \boldsymbol{x}) = p(\boldsymbol{x}|\boldsymbol{\theta}) \tilde{p}(\boldsymbol{\theta})$, which are then used to approximate the posterior given the observed data $\boldsymbol{x}_o$. 

Specifically, in this work we draw parameters $\boldsymbol{\theta} = \{\bd, \bdd, \s_\varepsilon\}$ from uniform priors (given in \refsec{priors}) and simulate the galaxy density field $\d_g$ using the forward model in \refeq{trivial_model} with $\LF$. The validity of $\LF$ as a robust and efficient simulator for SBI to infer the cosmological parameter $\s_8$ has been demonstrated in \cite{Tucci_2024} on mock data and in \cite{Nguyen_2024} for dark-matter halos. For each realization, we measure either the OC combinations (defined in \refsec{OCvsPB}) or the power spectrum P and bispectrum B (defined in \refsec{PB}).  

The OC data vector has $n \times N_\mathrm{bin}$ elements, where $N_\mathrm{bin}$ is the number of linear $k$-bins, and $n = 2,3,4$, or $6$, depending on the OC combination. The power spectrum plus bispectrum (P${+}$B) data vector has length $N_\mathrm{bin} + N_\mathrm{triangle}$, where $N_\mathrm{triangle}$ is the number of triangle configurations for the bispectrum.  

Our training set therefore consists of $N_\mathrm{sim}$ samples drawn from the joint distribution $\{\theta, \mathrm{OC}(\theta)\}$. Any finite training set is one particular random draw from the simulator: a different seed yields a different set of $\{\theta, \mathrm{OC}(\theta)\}$ pairs, leading to a slightly different trained posterior. To suppress this stochasticity, for each summary and each $\bdd^\mathrm{fid}$ we generate multiple independent realizations of $6\times10^4$ simulations with $\LF$, then concatenate and shuffle them into a single, larger training set, which we use throughout the paper. After training, we sample the estimated posterior $\mathcal{P}(\theta\,|\,\mathrm{OC}_o)$ conditioned on the observed data vector $\mathrm{OC}_o$, which is measured from a single $\LF$ realization of the galaxy field generated at the ground-truth parameter values. The observed data vector is held fixed across all independent seeds, so that the concatenation averages only over the stochasticity of the training simulations and not over the inference target itself. An analogous procedure is applied for the P${+}$B case.

Following \cite{Tucci_2024}, the SBI algorithm employed in this paper is as follows. For density estimation we use the neural posterior estimation method (NPE) \cite{Greenberg_2019} from the $\texttt{sbi}$ package \cite{Tejero-Cantero_2020} with normalizing flows, which model complex probability distributions via a series of invertible transformations. 

As a baseline configuration we initially adopted that of \cite{Tucci_2024}: a Masked Autoregressive Flow (MAF) \cite{Papamakarios_2017} NPE with 5 autoregressive layers, each a neural network with 2 fully connected layers of 50 hidden units and tanh activations, trained with a learning rate of $5\times10^{-4}$ and batch size 50. We found, however, that no fixed architecture performed consistently well across our five summary statistics, yielding posteriors whose width varied with the choice of network rather than with the information content of the summary itself. We therefore perform an independent hyperparameter optimization for each summary statistic, selecting per-summary the NPE configuration that minimizes the validation loss at fixed simulation budget. The optimization is carried out at three fiducial values $\bdd^\mathrm{fid}\in\{0,\,0.5,\,1\}$ per summary. The process, the choice of simulation budget, and the validation of the resulting posteriors are described in \refapp{SBI_validation}.

\subsection{Fisher information of summary statistics}
The Fisher information approximates the data vector likelihood as a Gaussian around the model prediction, and the posterior as a Gaussian around the best-fit parameters. In this way it provides a lower bound on expected parameter uncertainties. It serves as a useful, computationally inexpensive comparison point for the SBI results.

As defined in Chapter 14 of \cite{2020MCbook}, the Fisher matrix is given by
\begin{align}
F_{ij} \equiv - \left\langle \frac{\partial^2 \ln \mathcal{L}}{\partial \theta_i \partial \theta_j} \right\rangle \Bigg\rvert_{\hat{\boldsymbol{\theta}}},
\label{eq:fisher_matrix}
\end{align}
where $\hat{\boldsymbol{\theta}}$ is the maximum likelihood value of the parameters. In the linear-Gaussian approximation, for a data vector $\boldsymbol{x}(\boldsymbol{\theta})$ this reduces to:
\begin{align}
F_{ij} = \sum_{m,n} \frac{\partial x_m}{\partial \theta_i} \, (\mathrm{Cov}(\boldsymbol{x}))^{-1}_{mn} \, \frac{\partial x_n}{\partial \theta_j} \Bigg\rvert_{\hat{\boldsymbol{\theta}}},
\label{eq:fisher_matrix_gaussian}
\end{align}
and its inverse approximates the parameter covariance, $\mathrm{Cov}(\boldsymbol{\theta})\approx F^{-1}$. 
We evaluate \refeq{fisher_matrix_gaussian} numerically for each mock at its fiducial parameter values for all summary statistics. The derivatives $\partial x_m/\partial \theta_i$ are computed by finite differences with step sizes $\Delta\bd = \Delta\bdd = 0.1$ and $\Delta\s_\varepsilon = 0.05$, with the mean data vector at each step averaged over $N_\mathrm{der}$ realizations. The data covariance $\mathrm{Cov}(\boldsymbol{x})$ is estimated as the sample covariance of $N_\mathrm{cov}$ realizations at the fiducial parameters. The number of realizations is chosen according to the data-vector dimensionality of each summary (given in \refapp{SBI_validation}): we use $(N_\mathrm{der}, N_\mathrm{cov}) = (2000,\,4\times10^4)$ for P${+}$B, the highest-dimensional summary, and $(1000,\, 10^4)$ and $(2000,\,2\times10^4)$ for the second- and third-order OC summaries respectively. These choices were validated through convergence tests in which we varied the finite-difference step sizes as well as $N_\mathrm{der}$ and $N_\mathrm{cov}$, finding negligible impact on the Fisher matrices and marginalized parameter uncertainties over the adopted ranges.

An approximate result for the Fisher information at the field-level can be obtained from the MAP calculation (\refapp{MAP}; see also \cite{schmidt2025connection,pietroni/schmidt} for more general analytical treatments), which assumes a fiducial value of $\bdd=0$. This, however, does not include the stochastic amplitude contribution. When approximately including the noise by weighting each mode with its signal-to-noise squared, we obtain a result very close to the numerical FLI posterior for $\bdd=0$.

\subsection{Priors}\label{sec:priors}
For both FLI and SBI we adopt uniform box priors centered around the fiducial parameter values used in the mock generation and listed in \refsec{mocks}.

Specifically, for field-level inference we adopt uniform priors given by:
\begin{equation}
P^{\mathrm{FLI}}(\theta_i)
=
\mathcal{U}\!\left(
\theta_i^{\mathrm{fid}} - 0.5,\,
\theta_i^{\mathrm{fid}} + 0.5
\right),
\end{equation}
where $\theta_i \in \{\bd,\, \bdd,\, \s_\varepsilon\}$. 
For FLI the prior only needs to be wide enough not to dominate the posterior. For SBI the prior also serves as the distribution from which training simulations are drawn, so a wider prior spreads a fixed simulation budget over a larger volume, leaving fewer simulations near the posterior. To balance these requirements, for SBI we set prior bounds adaptively from Fisher forecasts. More specifically, for each fiducial value of $\bdd^{\mathrm{fid}}$, we use the corresponding Fisher-estimated standard deviations for the parameters $(\bd,\, \bdd,\, \s_\varepsilon)$ and define the prior bounds as
\begin{equation}
P^{\mathrm{SBI}}(\theta_i)
=
\mathcal{U}\!\left(
\theta_i^{\mathrm{fid}} - 10\,\mathrm{std\_dev}_{\mathrm{Fisher}}(\theta_i),\,
\theta_i^{\mathrm{fid}} + 10\,\mathrm{std\_dev}_{\mathrm{Fisher}}(\theta_i)
\right).
\end{equation}
The factor of 10 ensures the bounds comfortably enclose the posterior while keeping simulations concentrated where they are informative.

\section{Results}\label{sec:results}
In this section, we present inference results obtained with both field-level inference (FLI) and summary statistics, via simulation-based inference (SBI), on the mocks defined in \refsec{mocks}, for varying values of $\bdd^\mathrm{fid}$. Increasing $\bdd$ effectively introduces more nonlinearities in the modeled galaxy density field, allowing us to probe the information content beyond the linear power spectrum and bispectrum.

In order to provide a rough reference point for expected fiducial values $\bdd$, we recall the linear bias--power spectrum amplitude degeneracy for biased tracers, which is perfect at linear order. At second and higher order, this degeneracy is broken by advection (displacement) contributions, which do not introduce additional free coefficients (e.g., \cite{Sefusatti_2006,Desjacques_2018}). Instead, their coefficient is the linear bias, typically of order unity for galaxy samples. Thus, a typical amplitude to consider for the parameter $\bdd$ would be of order unity. We emphasize again, however, that due to the different forward model, parameter vector, and dataset employed, the results cannot be translated one-to-one to cosmological parameter inference. Instead, they illustrate in principle how FLI can extract more information from a weakly non-Gaussian field than low-order $n$-point functions while still in the perturbative regime.

\begin{figure}[tbp]
\centering 
\includegraphics[width=.85\textwidth]{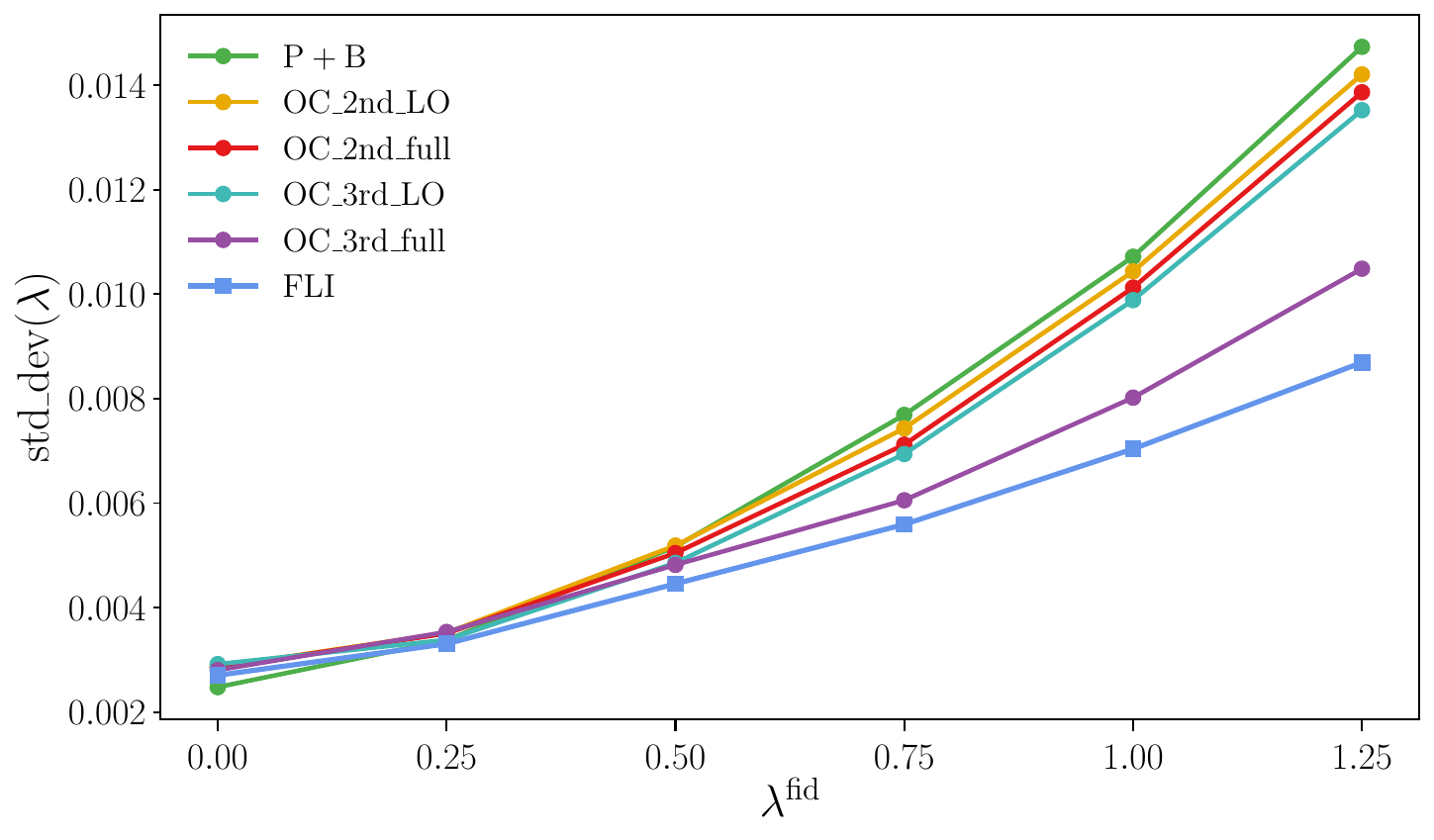}
\caption{\label{fig:OCs_vs_field} Inferred 68\% CL uncertainty on $\bdd$ as a function of $\bdd^\mathrm{fid}$, for field-level inference (blue) and for simulation-based inference with P${+}$B and the four OC combinations (colored as in the legend).}
\end{figure}

We begin by considering the case where all three parameters, $\bd$, $\bdd$, and $\s_\varepsilon$, are sampled simultaneously. Our analysis focuses on the posterior for $\bdd$ from the FLI and SBI posteriors, as this is the parameter that is constrained at leading order by the 3-point function (or OC$\_$2nd$\_$LO operator correlator). As shown in \reffig{OCs_vs_field}, the uncertainty on $\bdd$ increases with $\bdd^\mathrm{fid}$ for all methods, but substantially more slowly for FLI than for summary-based analyses. This is expected: for small $\bdd^\mathrm{fid}$, where the field remains close to linear, all methods agree to within the seed-to-seed scatter, and the low-order summaries capture nearly all of the available information, consistent with perturbative predictions. As nonlinearity grows (i.e. $\bdd^\mathrm{fid}$ grows), the low-order summaries fail to capture all the information encoded in higher-order correlations of the field, leading to larger uncertainties in the summary-based results. Among the summaries themselves, OC$\_$3rd$\_$full is visibly closer to FLI than the other four, which cluster together; we discuss this hierarchy in detail in \refsec{OC2ndvsOC3rd}.

\subsection{P+B vs OC}
\begin{figure}[tbp]
\centering 
\includegraphics[width=.85\textwidth]{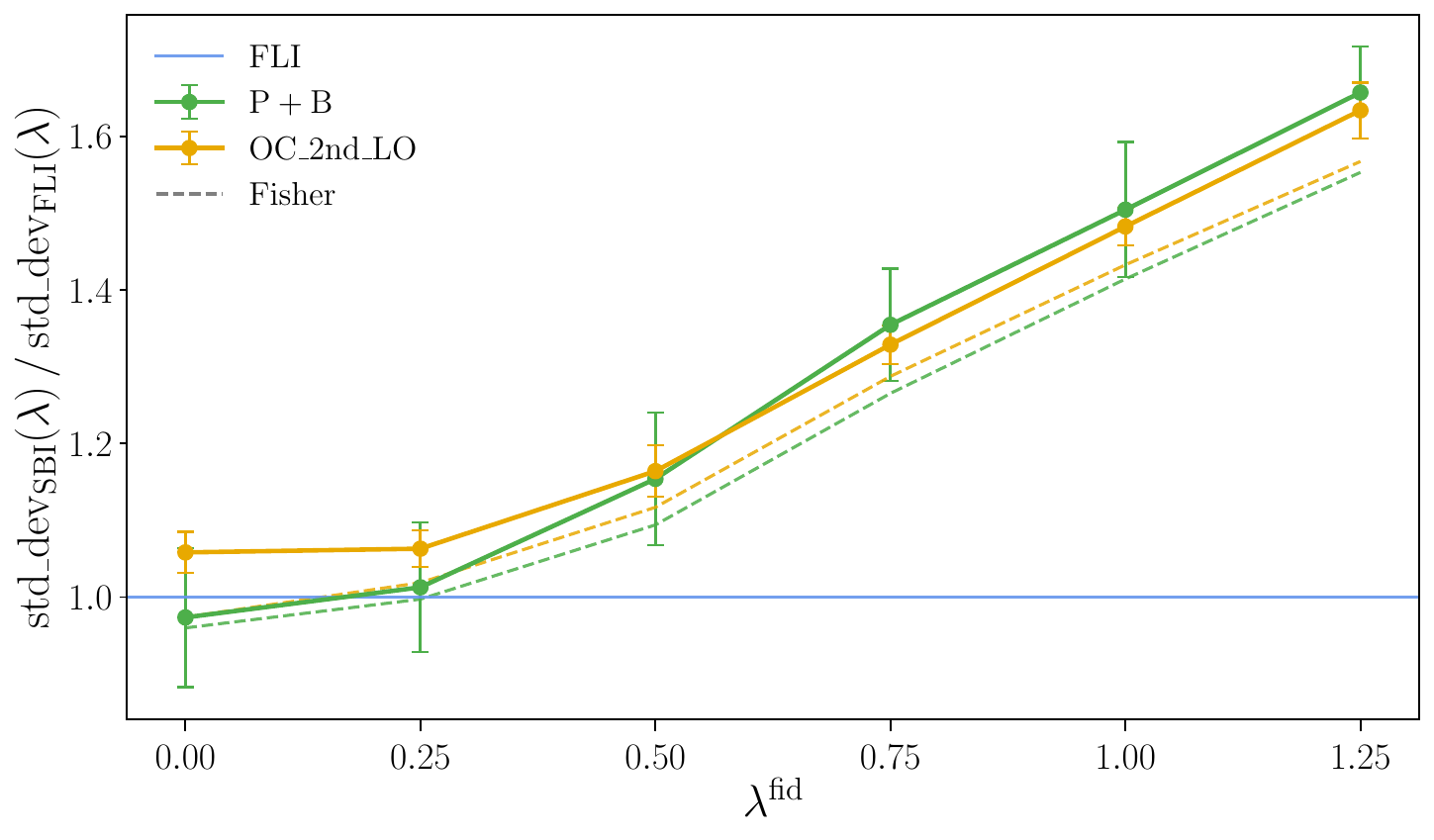}
\caption{\label{fig:PB_vs_OC2ndLO} Inferred 68\% CL uncertainty on $\bdd$ obtained with SBI, normalized by the FLI results, as a function of $\bdd^\mathrm{fid}$, focusing on P${+}$B (green)  and OC$\_$2nd$\_$LO (yellow) summaries. Dashed lines show the corresponding Fisher predictions. The line at unity marks the field-level result. Error bars indicate the scatter across independent seed realizations of the NPE network weights on the same training set.} 
\end{figure}

To understand how different summary statistics capture the information on $\bdd$, we first compare the OC$\_$2nd$\_$LO to the standard P${+}$B summaries. By construction (see \refsec{OCvsPB}), these two sets of summaries contain the same information at leading order in $\bdd$ in our setup. As shown in \reffig{PB_vs_OC2ndLO}, the SBI results for OC$\_$2nd$\_$LO (yellow) and P${+}$B (green) agree to within their error bars across most of the range of $\bdd^\mathrm{fid}$, in line with their expected equivalence at leading order. The only significant difference is at $\bdd^\mathrm{fid}=0$, as we discuss next.

For the considered range of $\bdd^\mathrm{fid}$, both SBI uncertainties lie slightly above the corresponding Fisher predictions (dashed lines), as expected since Fisher is a lower bound. The offset is comparable to the seed-to-seed scatter (defined in \refsec{methods}; see \refapp{SBI_validation} for the full breakdown across parameters), so SBI recovers the Fisher bound to within the precision of the method.

At $\bdd^\mathrm{fid}=0$ the P${+}$B and OC$\_$2nd$\_$LO summaries should be equivalent, and both should match the field-level constraint, since in the Gaussian limit all the information on $\bdd$ is contained in P${+}$B, or equivalently OC$\_$2nd$\_$LO. In the plot this holds only within the seed scatter: OC$\_$2nd$\_$LO sits marginally above the FLI line, while P${+}$B sits marginally below it, with the largest seed-to-seed scatter of any point in the comparison; both are consistent with unity within the estimated error. A possible explanation for the difficulty in obtaining reliable SBI posteriors in the $\bdd^{\rm fid}=0$ case is that the bispectrum mean vanishes in the Gaussian limit, so the ${\sim}\,925$ bispectrum entries of the P${+}$B data vector are dominated by sample fluctuations rather than signal, making the NPE training noisier.
Note moreover that the absolute uncertainty on $\bdd$ is smallest  at $\bdd^\mathrm{fid}=0$ (see \reffig{OCs_vs_field}, ${\sim}\,2\times10^{-3}$), so the precision required of the trained NPE is highest here; small training-stochasticity fluctuations therefore translate into the largest relative deviations.

\begin{figure}[tbp]
\centering 
\includegraphics[width=.49\textwidth]{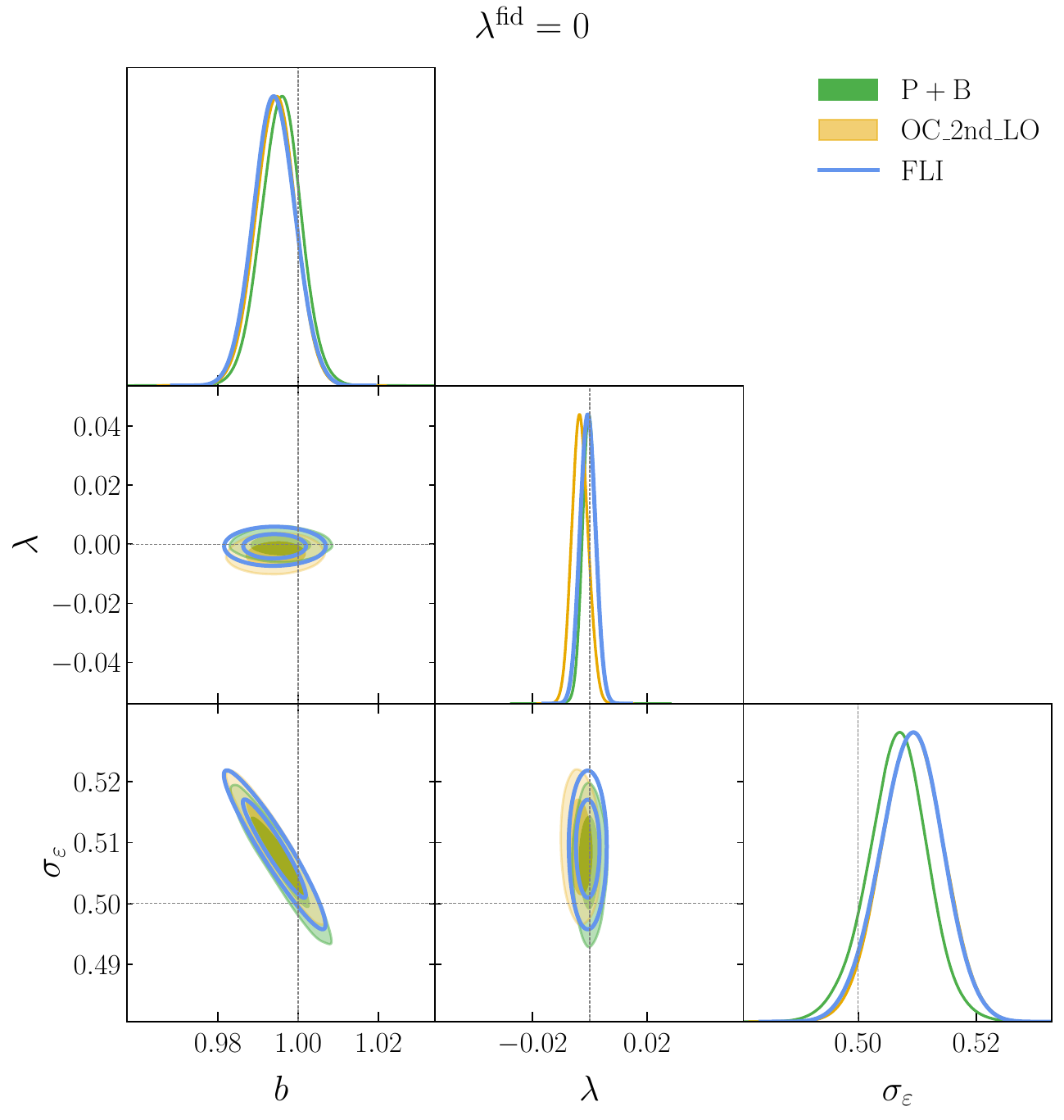}
\hfill
\includegraphics[width=.49\textwidth]{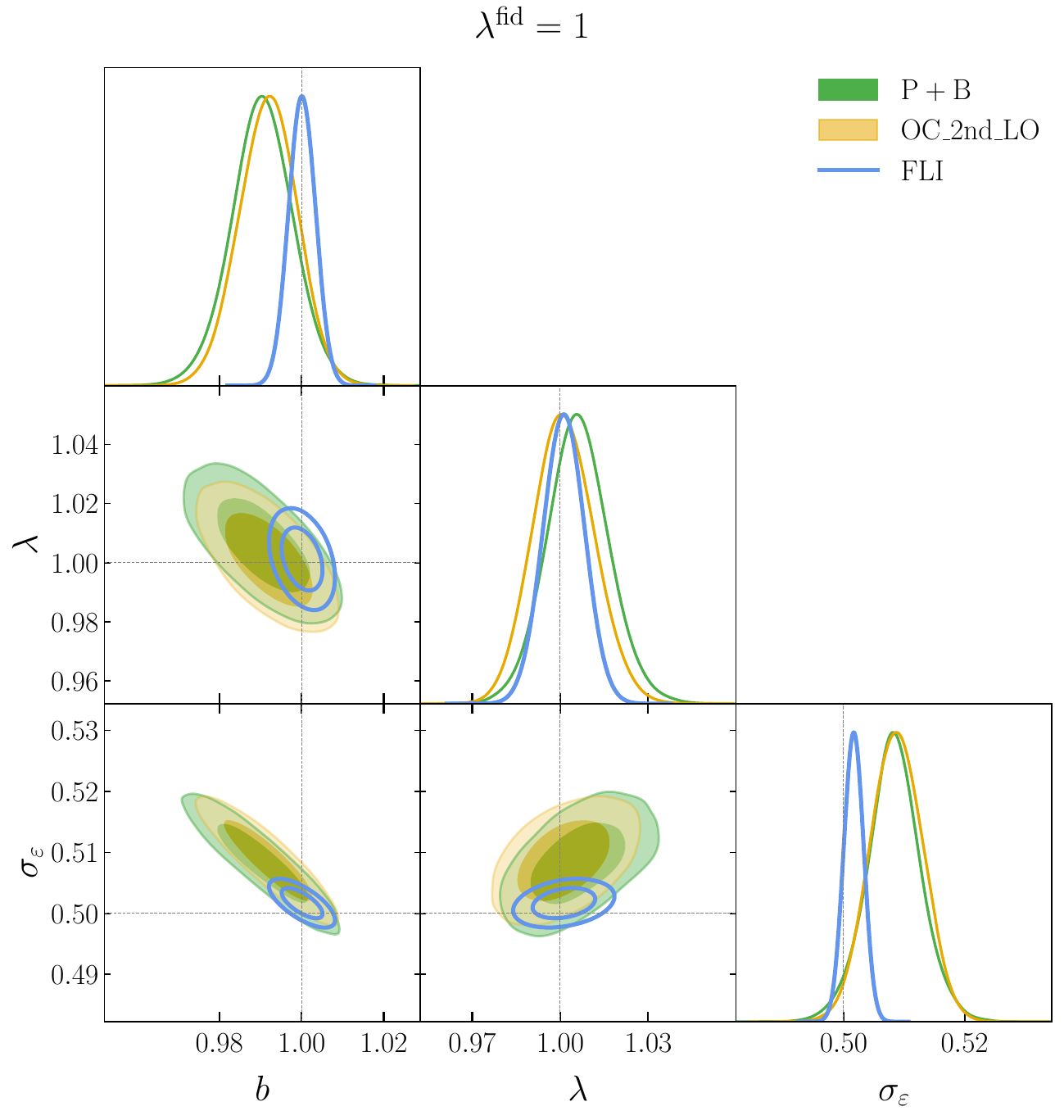}
\caption{\label{fig:posteriors_PB}  Parameter posteriors for $\bdd^\mathrm{fid}=0$ \emph{(left)} and $\bdd^\mathrm{fid}=1$ \emph{(right)}. Blue contours show the FLI posterior, green and yellow the SBI results for P${+}$B and OC$\_$2nd$\_$LO, respectively. 
Dashed lines mark the fiducial parameter values. The two panels use the same axis ranges for each parameter.}
\end{figure}

The full posteriors for the summaries and the field-level inference, shown in \reffig{posteriors_PB}, provide a complementary view of the trend observed in \reffig{OCs_vs_field}. The corner-plot posteriors are the pooled samples across NPE seeds (see \refapp{SBI_validation}). In the left panel, where $\bdd^\mathrm{fid} = 0$, all three posteriors have essentially the same width, consistent with the Gaussian-limit equivalence of P${+}$B, OC$\_$2nd$\_$LO and the field-level constraint. The posterior peaks for the three methods are slightly offset from one another, most visibly for $\s_\varepsilon$, where the FLI peak sits at a slightly higher value than the two summary peaks, but these shifts are consistent with the seed-to-seed and overall normalization scatter discussed above. 
In the right panel, for a finite $\bdd^\mathrm{fid}$, where significant nonlinearities are present in the data realization, the FLI posterior (blue) is noticeably tighter than both P${+}$B and OC$\_$2nd$\_$LO, highlighting the additional information captured by the full field. The gain is in fact most pronounced in $\bd$ and $\s_\varepsilon$.

The correlation between the $\bd$ and $\bdd$ posteriors, absent for $\bdd^\mathrm{fid}=0$ and anticorrelated for $\bdd^\mathrm{fid}>0$, matches the analytical expectation derived in \refapp{MAP} (strictly valid for small $\bdd$ and vanishing noise). The SBI summary posteriors also show a degeneracy between $\s_\varepsilon$ and $\bd$, present at all values of $\bdd^\mathrm{fid}$ and most pronounced at $\bdd^\mathrm{fid}=0$. The degeneracy direction corresponds to the overall clustering amplitude (variance) of the data, which is the best-measured combination from a low-order summary; breaking it requires either information on the scale-dependence of the clustering or, at $\bdd^\mathrm{fid}>0$, higher-point information about the field. The latter is accessed by FLI, which is why FLI gives a visibly tighter constraint on $\s_\varepsilon$ at finite $\bdd^\mathrm{fid}$ while the summaries remain broad.

\subsection{OC\_2nd vs OC\_3rd}\label{sec:OC2ndvsOC3rd}
\begin{figure}[tbp]
\centering 
\includegraphics[width=.85\textwidth]{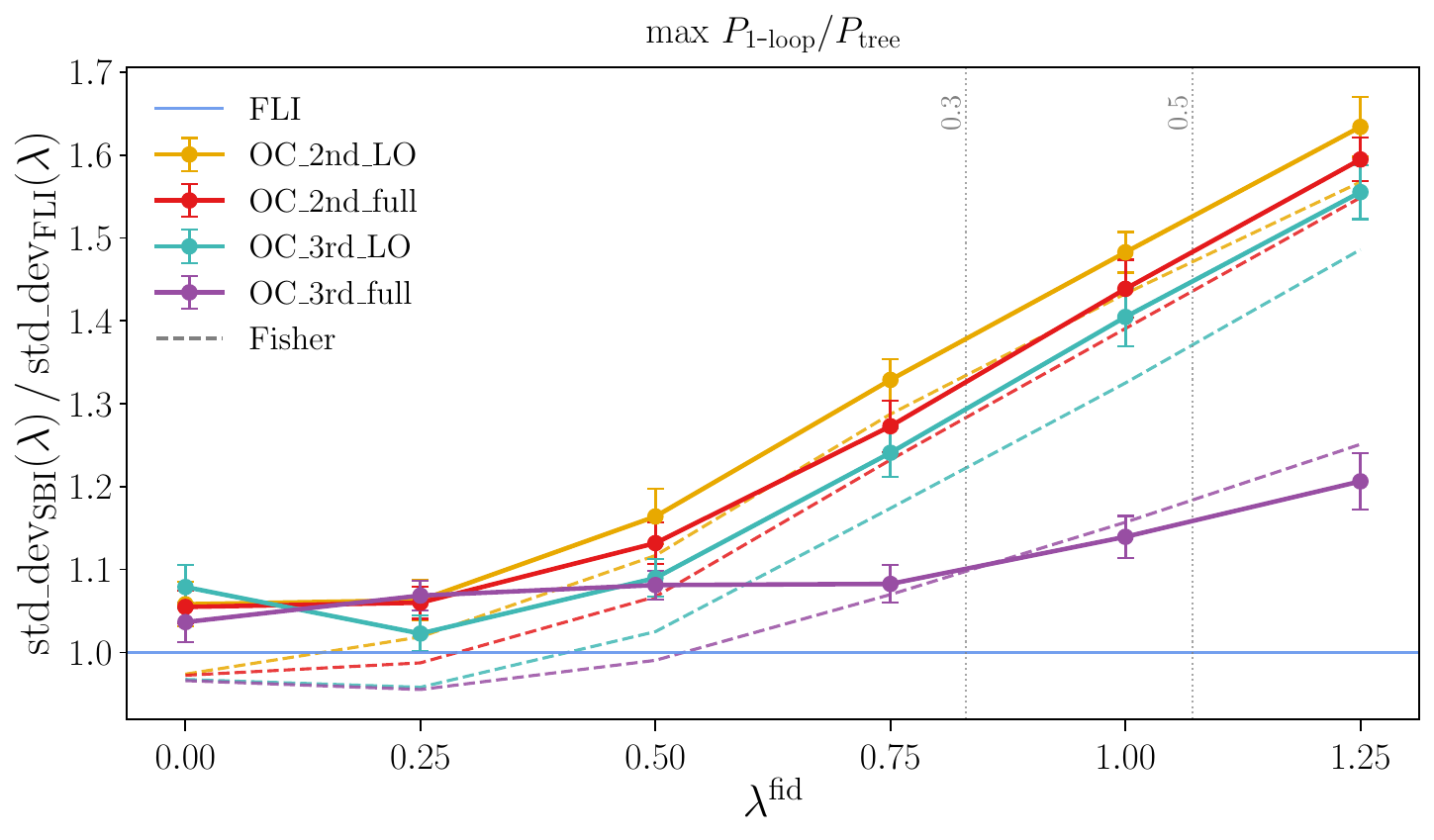}
\caption{\label{fig:OC_cases} Inferred 68\% CL error bar on $\bdd$ obtained with SBI using OC summaries, normalized by the FLI results, as a function of $\bdd^\mathrm{fid}$. Dashed lines show the Fisher predictions for each OC combination. Vertical dotted lines mark the values of $\bdd^\mathrm{fid}$  where the nonlinearity ratio \refeq{nl_ratio} equals $0.3$ and $0.5$; see \refapp{linear_regime}.}
\end{figure}

Next, we investigate how including higher-order OC combinations, which, by construction (see \refsec{OCvsPB}), contain progressively more information from higher-order correlations of the galaxy field, improves constraints on $\bdd$. \reffig{OC_cases} shows the 68\% CL posterior error on $\bdd$ obtained with SBI for different OC combinations, normalized by the corresponding FLI results, as a function of $\bdd^\mathrm{fid}$. 

The general trend is consistent with previous observations: the uncertainty on $\bdd$ relative to the FLI result grows with $\bdd^\mathrm{fid}$ for all OC combinations. Both Fisher (dashed) and SBI (solid) display the same hierarchy among the four summaries: as higher-order correlators are added at a given finite $\bdd^\mathrm{fid}$, the ratio decreases, showing that a larger fraction of the field-level information is captured.

While the comparison itself remains valid for any value of $\bdd$ in the context of the specific model \refeq{trivial_model}, the regime of validity of \refeq{trivial_model} in its EFT interpretation depends on $\bdd^\mathrm{fid}$, as quantified by the nonlinearity ratio $R(\bdd)$ defined in \refeq{nl_ratio}. The vertical dotted lines in \reffig{OC_cases} mark $\bdd^\mathrm{fid}=0.83$ and $\bdd^\mathrm{fid}=1.07$, where $R=0.3$ and $R=0.5$ respectively; see \refapp{linear_regime} for the explicit dependence.

Focusing now on the perturbative range, we quantify the relative gain of FLI over higher-order summaries. At $\bdd^\mathrm{fid}=0$ the information on $\bdd$ is fully captured by P${+}$B or, equivalently, OC$\_$2nd$\_$LO: higher-order OCs of a Gaussian field add no new information on $\bdd$ beyond what is already contained in the cross-correlator $\langle\d_g\,\d_g^2\rangle$. Consistent with this, the normalized OC ratios scatter around unity at the few-percent level, within the seed uncertainty. 

At $\bdd^\mathrm{fid}=0.75$ the nonlinearity ratio \refeq{nl_ratio} equals $R \approx 0.23$, well within the perturbative regime. The current state of the art in large-scale-structure analyses corresponds to P${+}$B, equivalent here to OC$\_$2nd$\_$LO, whose SBI uncertainty on $\bdd$ is $\sim 35\%$ larger than the field-level result at this point; the Fisher forecast gives $\sim 30\%$. Adding higher-order composite-operator correlators reduces this gap. Quantitatively, in Fisher the uncertainties for OC$\_$2nd$\_$full, OC$\_$3rd$\_$LO and OC$\_$3rd$\_$full sit at $\sim 25\%$, $\sim 15\%$ and $\sim 5\%$ above FLI; in SBI they sit at $\sim 30\%$, $\sim 20\%$ and $\sim 10\%$. A similar pattern holds at $\bdd^\mathrm{fid}=1$, where $R\approx 0.43$: the Fisher values for OC$\_$2nd$\_$LO through OC$\_$3rd$\_$full are $\sim 45\%$, $\sim 40\%$, $\sim 35\%$ and $\sim 15\%$, while the SBI values are $\sim 50\%$, $\sim 45\%$, $\sim 35\%$ and $\sim 15\%$. The two methods thus agree closely on the cubic-order summaries OC$\_$3rd$\_$LO and OC$\_$3rd$\_$full, while the SBI ratios sit slightly above Fisher for the lower-order summaries; given the seed-to-seed precision floor of the trained NPE (cf.\ error bars in \reffig{OC_cases}), the Fisher values provide the more reliable indication of how the information is distributed among these summaries.
We note that for OC$\_$3rd$\_$full the SBI uncertainty falls marginally below the corresponding Fisher prediction at $\bdd^\mathrm{fid}\gtrsim 1$; the offset is comparable to the seed-to-seed scatter of the trained NPE (cf.\ error bars in \reffig{OC_cases}) and is consistent with statistical fluctuations of the inference pipeline rather than a genuine sub-Fisher constraint. Overall, the bulk of the information accessible to summary statistics beyond P${+}$B is captured only once the full cubic correlator is included, i.e. resides in the higher $n$-point functions, yet even OC$\_$3rd$\_$full does not fully reach the field-level constraint. 

\begin{figure}[tbp]
\centering 
\includegraphics[width=.85\textwidth]{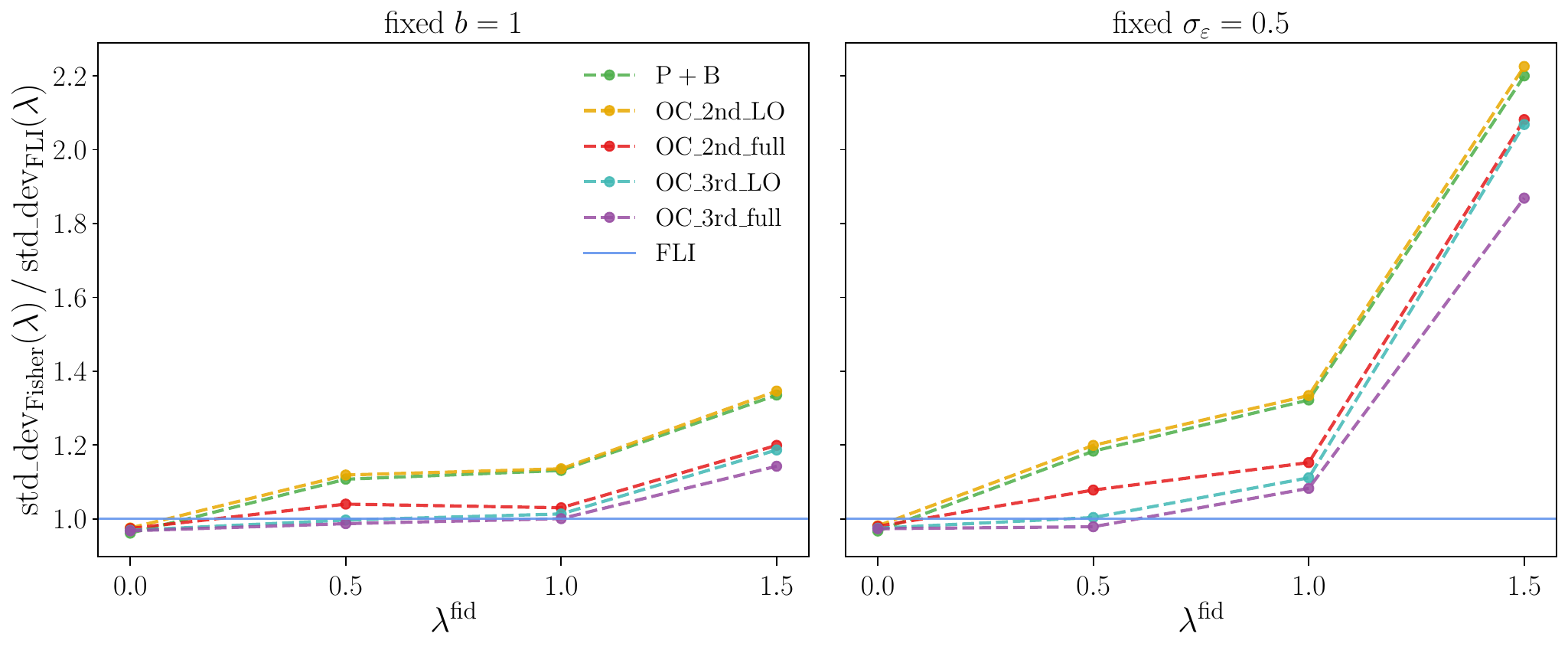}
\caption{\label{fig:fixed_params} Fisher forecast for the uncertainty on $\bdd$, normalized by the FLI result, as a function of $\bdd^\mathrm{fid}$, for fixed linear bias $\bd=1$ \emph{(left)} and fixed stochastic amplitude $\s_\varepsilon=0.5$ \emph{(right)}. Each curve corresponds to one summary, as labeled.}
\end{figure}

The full posteriors shown in \reffig{posteriors_PB} revealed noticeable degeneracies between parameters. To explore how these impact parameter constraints, we examine the effect of fixing individual parameters on the normalized 68\% CL error bar on $\bdd$. 
\reffig{fixed_params} shows this for two cases: fixing the linear bias $\bd=1$ \textit{(left)} and fixing the stochastic amplitude $\s_\varepsilon=0.5$ \textit{(right)}. For this comparison we extend the range of $\bdd^\mathrm{fid}$ to $1.5$, slightly beyond the $1.25$ used in the main analysis, since the Fisher forecast is inexpensive to evaluate and the additional point helps to resolve the trend at high nonlinearity. Note that the rightmost points ($\bdd^\mathrm{fid}=1.5$) sit at $R\approx 1$, at the edge of the perturbative regime (see \refapp{linear_regime}).

The right panel, where the stochastic amplitude is fixed, behaves qualitatively like the all-parameters-free case (compare to \reffig{OCs_vs_field} and \reffig{OC_cases}): the summary-to-FLI ratios still rise steeply with $\bdd^\mathrm{fid}$. This indicates that the dominant source of the $\bdd$ uncertainty in the all-free case is not the stochastic amplitude, but the combination of $\bdd$ with $\bd$.

By contrast, the left panel, where the linear bias $\bd$ is fixed, shows significantly smaller ratios across the whole range of $\bdd^\mathrm{fid}$. For example, the normalized 68\% CL error bars on $\bdd$ for OC$\_$2nd$\_$LO and P${+}$B are consistent with FLI at $\bdd^\mathrm{fid}=0$ and rise to only $\sim 15\%$ above FLI at $\bdd^\mathrm{fid}=1$; for the higher-order summaries the rise is essentially negligible (within $\sim 5\%$ of FLI). This contrasts with the substantially larger values reached when $\bd$ is free, $\sim 35$--$45\%$ above FLI at $\bdd^\mathrm{fid}=1$ for all summaries except OC$\_$3rd$\_$full. This demonstrates that in the all-free case, part of the uncertainty on $\bdd$ arises from the partial degeneracy between $\bdd$ and $\bd$ seen in \reffig{posteriors_PB}. The summary statistics suffer more strongly from this degeneracy than FLI, which is why fixing $\bd$ narrows the gap between them.

\section{Conclusions}

In this work we have compared, in a controlled setting, the constraining power of field-level inference with that of a family of summary statistics: power spectrum plus bispectrum (P${+}$B), and a hierarchy of composite-operator correlators (OCs). The forward model is the minimal nonlinear model of \refeq{trivial_model}, with three sampled parameters in addition to the marginalization over the initial conditions: the linear bias $\bd$, the quadratic-bias parameter $\bdd$, and the stochastic amplitude $\s_\varepsilon$. FLI is carried out by jointly sampling the initial conditions and parameters via MCMC, while the summary posteriors are obtained with simulation-based inference (SBI) via neural posterior estimation. The composite-operator correlators are built from the same forward model that FLI samples, so that, at the level of the underlying physics, the two analyses differ only in the level of compression applied to the data.

Our main findings are as follows. First, in the Gaussian limit $\bdd^\mathrm{fid}=0$ all methods agree to within the training-stochasticity scatter, as expected analytically. Second, as $\bdd^\mathrm{fid}$ grows so does the gap to the field-level result: at $\bdd^\mathrm{fid}=0.75$, still safely within the perturbative regime by the criterion of \refeq{nl_ratio}, the standard P${+}$B summary (equivalent here to OC$\_$2nd$\_$LO) yields a $68\%$ CL uncertainty on $\bdd$ that is $\sim 35\%$ larger than the FLI result, rising to $\sim 50\%$ at $\bdd^\mathrm{fid}=1$. Third, extending the summary to higher-order composite-operator correlators reduces this gap; including the full set of up to cubic correlators, OC$\_$3rd$\_$full, reduces the information loss relative to FLI to $\sim 10\%$ at $\bdd^\mathrm{fid}=0.75$ and $\sim 15\%$ at $\bdd^\mathrm{fid}=1$. Even OC$\_$3rd$\_$full does not fully reach FLI in the range considered, showing that some information remains in correlators of even higher order than those captured by OC$\_$3rd$\_$full. It appears that, for this particular setup, the bulk of the FLI advantage is in breaking the $\bd$--$\bdd$ degeneracy: when $\bd$ is fixed, the FLI/summary gap shrinks substantially, while fixing $\s_\varepsilon$ has essentially no effect (\reffig{fixed_params}). 

These findings connect to the analogous structure in real-space biased-tracer analyses. In the context of $\s_8$ inference from rest-frame galaxy clustering, the role played here by $\bdd$ is taken by the coefficient of the second-order displacement, which in that setting is the linear bias $b_1$ itself \cite{Desjacques_2018}. In that case, additional higher-order operators also break the leading-order amplitude--bias degeneracy, mirroring the role of higher-order OC summaries here, via the subleading terms discussed below \refeq{b_2_map}. Our results therefore suggest that the FLI advantage observed in realistic cosmological forward models \cite{Andrews_2023,Nguyen_2024,Tucci_2024} is at least in part driven by this same mechanism: a more efficient use of the higher-order information that breaks bias--amplitude degeneracies in the perturbative regime.

Two additional studies are reported in the appendices. First, at lower noise ($\s^\mathrm{fid}_\varepsilon=0.3$, \refapp{lower_noise}) the gap between summaries and FLI widens for all three parameters and all four OC summaries: at $\bdd^\mathrm{fid}=1$, the Fisher forecast for the OC$\_$2nd$\_$LO uncertainty on $\bdd$ is $\sim 90\%$ above the FLI value at $\s^\mathrm{fid}_\varepsilon=0.3$, compared to $\sim 45\%$ above FLI at the fiducial $\s^\mathrm{fid}_\varepsilon=0.5$, and the same trend is visible for $\bd$ and (most pronouncedly) for $\s_\varepsilon$ itself. This is expected, as a higher signal-to-noise per mode affects higher-order statistics proportionally more strongly. This finding demonstrates that the order in $n$-point statistics necessary to saturate the field-level information also depends on the noise level; while OC$\_$3rd$\_$full was already close to optimal for the fiducial noise level, it no longer is at lower noise (cf. \reffig{lower_sigma}). This further motivates the use of full field-level (or higher-order OC) methods for upcoming high-density surveys.

Second, the OCs offer a substantial computational advantage over direct $n$-point estimation: their data-vector dimensionality grows polynomially with the maximum operator order, rather than exponentially with $n$ (\refsec{OCvsPB}, \refapp{SBI_validation}). In our configuration the largest OC data vector (120 entries for OC$\_$3rd$\_$full) is nearly an order of magnitude smaller than the P${+}$B vector (945 entries) and roughly two orders smaller than what including the trispectrum would require ($\sim 1.5\times 10^4$ entries). This makes OCs well suited for SBI, where simulation cost is often the dominant bottleneck.

On the methodological side, two findings about the SBI pipeline itself are worth noting. No fixed NPE architecture performs uniformly well across all five summaries: network hyperparameters optimized for P${+}$B are suboptimal for the OCs and vice versa, with the resulting spread in posterior widths exceeding the run-to-run seed scatter. We therefore optimize the hyperparameters separately for each summary (\refapp{SBI_validation}), and verify that the resulting production posteriors match the Fisher predictions to within the seed scatter for every parameter (\reffig{seed_validation}). Without per-summary tuning, the comparison between summaries would be contaminated by architecture mismatch rather than reflecting their actual information content. Even with per-summary tuning, the seed-to-seed scatter of the trained NPE sets a precision floor of $\sim 2$--$3\%$ on the marginal posterior width.

Several extensions are natural. The forward model used here is deliberately minimal; a realistic setting includes nonlinear gravitational evolution, additional bias operators (tidal, third-order) and non-Gaussian noise contributions, the latter potentially generating new OC shapes. While we have not explicitly tested non-Gaussian noise, fixing the (Gaussian) noise amplitude $\s_\varepsilon$ left the relative FLI/summary improvement essentially unchanged (\reffig{fixed_params}). We leave the corresponding study to future work.

\acknowledgments
We thank Ivana Babić, Noemi Anau Montel, Adri Duivenvoorden, Sam Goldstein, \c{S}afak \c{C}elik, Adrian Bayer, Julia Stadler, and Azadeh Moradinezhad Dizgah for valuable discussions during the preparation of this publication.

Computations were performed on the
\href{https://docs.mpcdf.mpg.de/doc/computing/clusters/systems/Astrophysics/MPA-FREYA.html}{FREYA} 
and 
\href{https://docs.mpcdf.mpg.de/doc/computing/clusters/systems/Astrophysics/MPA-ORION.html}{ORION} 
clusters, maintained by the 
\href{https://www.mpcdf.mpg.de}{Max Planck Computing \& Data Facility}.

\appendix

\section{Details of the MAP calculation}\label{appendix:MAP}
We here provide more details on the derivation of the MAP point for $\bdd$ from \refsec{zero_noise}, including details of the inverse solution in \refeq{inverse_solution} and the Jacobian in terms of the forward model around this solution. 

The two solutions to the constraint $\d_{g,\mathrm{det}}[\d]=\d_g$ imposed by the Dirac delta in \refeq{posterior_zero} are
\begin{align}
\d_{g,\mathrm{det}}^{-1}[\d_g, \bd, \bdd](\boldsymbol{x}) = \frac{-\bd \pm \sqrt{\bd^2+4\bdd\d_g(\boldsymbol{x})}}{2\bdd}.
\label{eq:two_solution_appendix}
\end{align}
If $\d_g$ is small enough or $\bdd$ is not too large, the discriminant is non-negative and both solutions are real. For $\bd>0$, we choose the $+$ branch that connects smoothly to the linear solution in the limit $\bdd\to 0$, when the model reduces to $\d_g\approx \bd \d$. Taylor-expanding
\begin{align}
\d_{g,\mathrm{det}}^{-1}[\d_g, \bd, \bdd](\boldsymbol{x}) & = \frac{1}{2}\bdd^{-1}\left[\sqrt{\bd^2+4\bdd\d_g(\boldsymbol{x})}-\bd\right]\\ 
& = \frac{1}{2}\bdd^{-1}\left[\bd+2\frac{\bdd}{\bd}\d_g(\boldsymbol{x})-2\frac{\bdd^2}{\bd^3}\d_g^2(\boldsymbol{x})+4\frac{\bdd^3}{\bd^5}\d_g^3(\boldsymbol{x})-\bd\right]+\mathcal{O}(\d_g^4) \nonumber \\
& = \frac{1}{\bd}\d_g(\boldsymbol{x})-\frac{\bdd}{\bd^3}\d_g^2(\boldsymbol{x})+2\frac{\bdd^2}{\bd^5}\d_g^3(\boldsymbol{x})+\mathcal{O}(\d_g^4).
\label{eq:ix_appendix}
\end{align}
In the main analysis in \refsec{zero_noise} we truncate at $O(\d_g^2)$, i.e. at linear order in $\bdd$.

It remains to express the Jacobian in terms of the forward model evaluated at $\d=\d_{g,\mathrm{det}}^{-1}$. From \refeq{trivial_model}, the functional derivative is diagonal in real space
\begin{equation}
\frac{\mathcal{D}\d_{g,\mathrm{det}}(\boldsymbol{x})}{\mathcal{D}\d(\boldsymbol{y})} 
= \left[\bd + 2\bdd\,\d(\boldsymbol{x})\right] \d_\mathrm{D}(\boldsymbol{x}-\boldsymbol{y})\,.
\label{eq:func_deriv}
\end{equation}
The corresponding determinant and its logarithm are therefore
\begin{align}
\left|\frac{\mathcal{D}\d_{g,\mathrm{det}}}{\mathcal{D}\d}\right| 
&= \prod_{\boldsymbol{x}} \left|\bd + 2\bdd\,\d(\boldsymbol{x})\right|\,, 
\label{eq:Jacobian}\\
\ln\left|\frac{\mathcal{D}\d_{g,\mathrm{det}}}{\mathcal{D}\d}\right| 
&= \int_{\boldsymbol{x}} \ln\left|\bd + 2\bdd\,\d(\boldsymbol{x})\right|\,.
\label{eq:Jacobian_ln}
\end{align} 
Inserting \refeq{Jacobian_ln} and \refeq{inverse_solution} into 
\refeq{posterior_integrated} and expanding to $\mathcal{O}(\d_g^2)$ in the inverse solution gives
\begin{align}
-2\ln \mathcal{P}[\bd, \bdd|\d_g] 
\overset{\mathcal{O}(\d_g^2)}{=}\ & \int_{\vk}^\Lambda\frac{1}{P_\mathrm{L}(k)}
\left|\frac{1}{\bd}\d_g(\vk) - \frac{\bdd}{\bd^3}(\d_g^2)(\vk)\right|^2 \nonumber \\
& + 2\int_{\boldsymbol{x}} \ln\left[\bd + 2\frac{\bdd}{\bd}\d_g(\boldsymbol{x}) 
- 2\frac{\bdd^2}{\bd^3}\d_g^2(\boldsymbol{x})\right]\,.
\label{eq:solution_in}
\end{align}
The first term is a Fourier-space integral over $\d_g$ and $(\d_g^2)$, while the second term, the Jacobian contribution, is a real-space integral over the same fields.

We can now evaluate the MAP point for $\bdd$, keeping $\bd$ fixed. Taking the derivative of \refeq{solution_in} with respect to $\bdd$:
\begin{align}
-2 \frac{\partial}{\partial \bdd}\ln\mathcal{P}[\bd, \bdd|\d_g] \overset{\mathcal{O}(\d_g^2)}{=}  & -\frac{2}{\bd^3}\int_{\vk}^\Lambda\frac{1}{P_\mathrm{L}(k)}(\d_g^2)(-\vk)\left[\frac{1}{\bd}\d_g(\vk)-\frac{\bdd}{\bd^3}(\d_g^2)(\vk)\right]\nonumber \\
& +4\int_{\boldsymbol{x}} \left[\bd+2\frac{\bdd}{\bd}\d_g(\boldsymbol{x})-2\frac{\bdd^2}{\bd^3}\d_g^2(\boldsymbol{x})\right]^{-1}\left[\frac{1}{\bd}\d_g(\boldsymbol{x})-2\frac{\bdd}{\bd^3}\d_g^2(\boldsymbol{x})\right]\,.
\label{eq:part_b2}
\end{align}
If we linearize the second part, again only keeping up to second order in $\d_g$, we get:
\begin{align}
-2 \frac{\partial}{\partial \bdd}\ln\mathcal{P}[\bd, \bdd|\d_g] \overset{\mathcal{O}(\d_g^2)}{=}  & -\frac{2}{\bd^3}\int_{\vk}^\Lambda\frac{1}{P_\mathrm{L}(k)}(\d_g^2)(-\vk)\left[\frac{1}{\bd}\d_g(\vk)-\frac{\bdd}{\bd^3}(\d_g^2)(\vk)\right]\nonumber \\
& +4\int_{\boldsymbol{x}} \left[\frac{1}{\bd^2}\d_g(\boldsymbol{x})-4\frac{\bdd}{\bd^4}\d_g^2(\boldsymbol{x})\right]\,.
\label{eq:part_b2_linear}
\end{align}
Setting this to zero leads to:
\begin{align}
 -2\frac{\bdd}{\bd^6}&\int_{\vk}^\Lambda\frac{1}{P_\mathrm{L}(k)}(\d_g^2)(-\vk)(\d_g^2)(\vk) + 16\frac{\bdd}{\bd^4}\int_{\boldsymbol{x}}\d_g^2(\boldsymbol{x})= \nonumber \\
 -\frac{2}{\bd^4}&\int_{\vk}^\Lambda\frac{1}{P_\mathrm{L}(k)}(\d_g^2)(-\vk)\d_g(\vk) + \frac{4}{\bd^2}\int_{\boldsymbol{x}} \d_g(\boldsymbol{x})\,.
\label{eq:part_b2_zero}
\end{align}
Multiplying with $-\bd^2/2$ yields
\begin{align}
\frac{\bdd}{\bd^4}\int_{\vk}^\Lambda\frac{1}{P_\mathrm{L}(k)}(\d_g^2)(-\vk)(\d_g^2)(\vk) - 8\frac{\bdd}{\bd^2}\int_{\boldsymbol{x}}\d_g^2(\boldsymbol{x})= \frac{1}{\bd^2}\int_{\vk}^\Lambda\frac{1}{P_\mathrm{L}(k)}(\d_g^2)(-\vk)\d_g(\vk) - 2\int_{\boldsymbol{x}} \d_g(\boldsymbol{x})\,.
\label{eq:part_b2_zero_clean}
\end{align}
Both real-space integrals of $\d_g$ in \refeq{part_b2_zero} vanish: $\int_{\boldsymbol{x}}\d_g(\boldsymbol{x}) = \d_g(\vk=0)=0$, since the $\vk=0$ mode is excluded throughout (see \refsec{zero_noise}). Dropping these terms, \refeq{part_b2_zero_clean} can be solved for $\bdd$, yielding the MAP estimate quoted in \refeq{b_2_map}: the numerator $N[\d_g]$ is the (surviving) right-hand side, and the denominator $D[\d_g]$ is the bracket multiplying $\bdd$ on the left-hand side.

We can use the same expansion to obtain the leading-order Fisher information by taking another derivative of \refeq{part_b2_linear} and the expectation value. We evaluate at the fiducial $\bdd=0$, using that the odd (three-point) term $\langle(\d_g^2)(-\vk)\d_g(\vk)\rangle$ and $\langle\int_{\boldsymbol{x}}\d_g(\boldsymbol{x})\rangle$ both vanish there:
\begin{align}
  - \left\langle\frac{\partial^2}{\partial \bdd^2}\ln\mathcal{P}[\bd, \bdd|\d_g]\right\rangle =
 \frac{1}{\bd^6}\int_{\vk}^\Lambda\frac{1}{P_\mathrm{L}(k)}\langle(\d_g^2)(-\vk)(\d_g^2)(\vk)\rangle - 8\frac{1}{\bd^4}\int_{\boldsymbol{x}} \langle\d_g^2(\boldsymbol{x})\rangle\,,
  \\
  - \left\langle\frac{\partial^2}{\partial \bdd \partial \bd}\ln\mathcal{P}[\bd, \bdd|\d_g]\right\rangle = \frac{2\bdd}{\bd^5} \left[
 - \frac{3}{\bd^2} \int_{\vk}^\Lambda\frac{1}{P_\mathrm{L}(k)}\langle(\d_g^2)(-\vk)(\d_g^2)(\vk)\rangle + 16\int_{\boldsymbol{x}}\langle\d_g^2(\boldsymbol{x})\rangle\right]\,.
\end{align}
The expectation value of the first expression yields the Fisher information on $\bdd$, $F_{\bdd\bdd}$, while the expectation value of the latter corresponds to $F_{\bdd\bd}$. Numerical evaluation shows that the second, positive term is larger, so that $F_{\bdd\bd} \geq 0$ for $\bdd \geq 0$. Thus, we find that the posterior covariance between $\bdd$ and $\bd$ scales as $-\bdd/\bd^3$.

\section{Perturbative regime and nonlinearity criterion}\label{appendix:linear_regime}

\begin{figure}[tbp]
\centering
\includegraphics[width=0.75\textwidth]{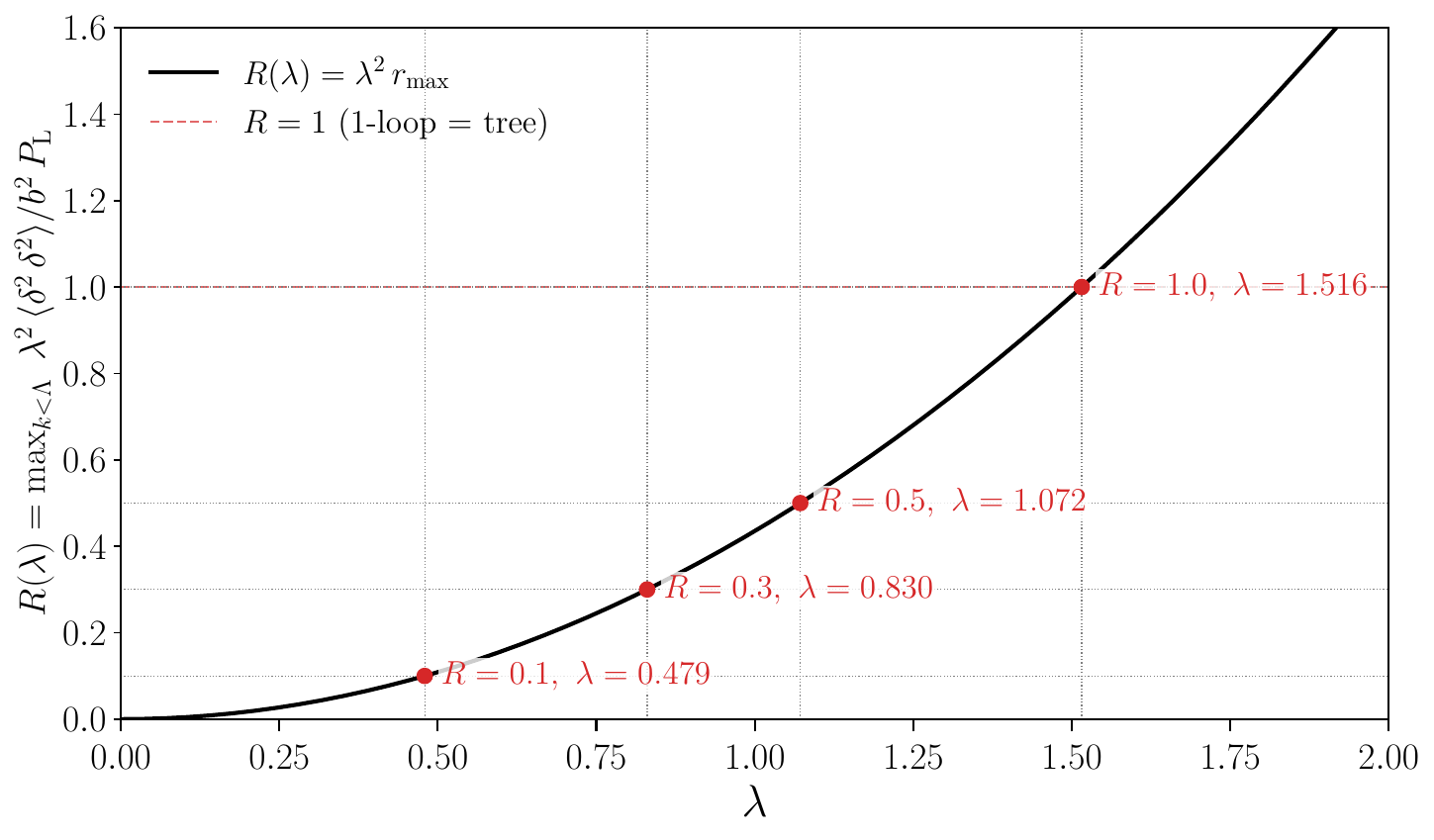}
\caption{\label{fig:linear_regime}
Nonlinearity ratio $R(\bdd)$ defined in \refeq{nl_ratio} as a function of $\bdd$, at $\bd=1$ and $k_\mathrm{max}\equiv\Lambda=0.14\,h\,\mathrm{Mpc}^{-1}$. The dashed red line marks $R=1$, the point where the 1-loop and tree-level contributions are equal. Red dots indicate $R \in \{0.1,\, 0.3,\, 0.5,\, 1.0\}$ with the corresponding $\bdd$ values labeled. The vertical dotted lines in \reffig{OC_cases} correspond to $R=0.3$ and $R=0.5$.}
\end{figure}

The nonlinearity ratio $R(\bdd)$ defined in \refeq{nl_ratio} measures how perturbative the forward model \refeq{trivial_model} is when interpreted as the lowest-order terms of an EFT expansion. \reffig{linear_regime} shows $R(\bdd)$ as a function of $\bdd$, evaluated at $\bd=1$ and $k_\mathrm{max}\equiv\Lambda=0.14\,h\,\mathrm{Mpc}^{-1}$. The condition $R=1$, marking the formal breakdown of perturbation theory, is reached at $\bdd \approx 1.52$. More conservative thresholds give $R=0.5$ at $\bdd \approx 1.07$ and $R=0.3$ at $\bdd \approx 0.83$; these two values are the ones marked as vertical dotted lines in \reffig{OC_cases}.

The fiducial values used in our main analysis span $\bdd^\mathrm{fid} \in \{0,\,0.25,\,0.5,\,0.75,\,1.0,\,1.25\}$. The first four satisfy $R<0.3$ and sit safely within the perturbative regime: $R(0.25)\approx 0.03$, $R(0.5)\approx 0.11$, $R(0.75)\approx 0.23$. At $\bdd^\mathrm{fid}=1.0$ we have $R\approx 0.43$, and at $\bdd^\mathrm{fid}=1.25$, $R\approx 0.68$, so the two largest fiducial values probe the transition region where higher-order corrections to the forward model start to become non-negligible.

\section{Posterior sampling and validation}
\subsection{FLI}\label{appendix:FLI_validation}
\begin{figure}[tbp]
\centering 
\includegraphics[width=.85\textwidth]{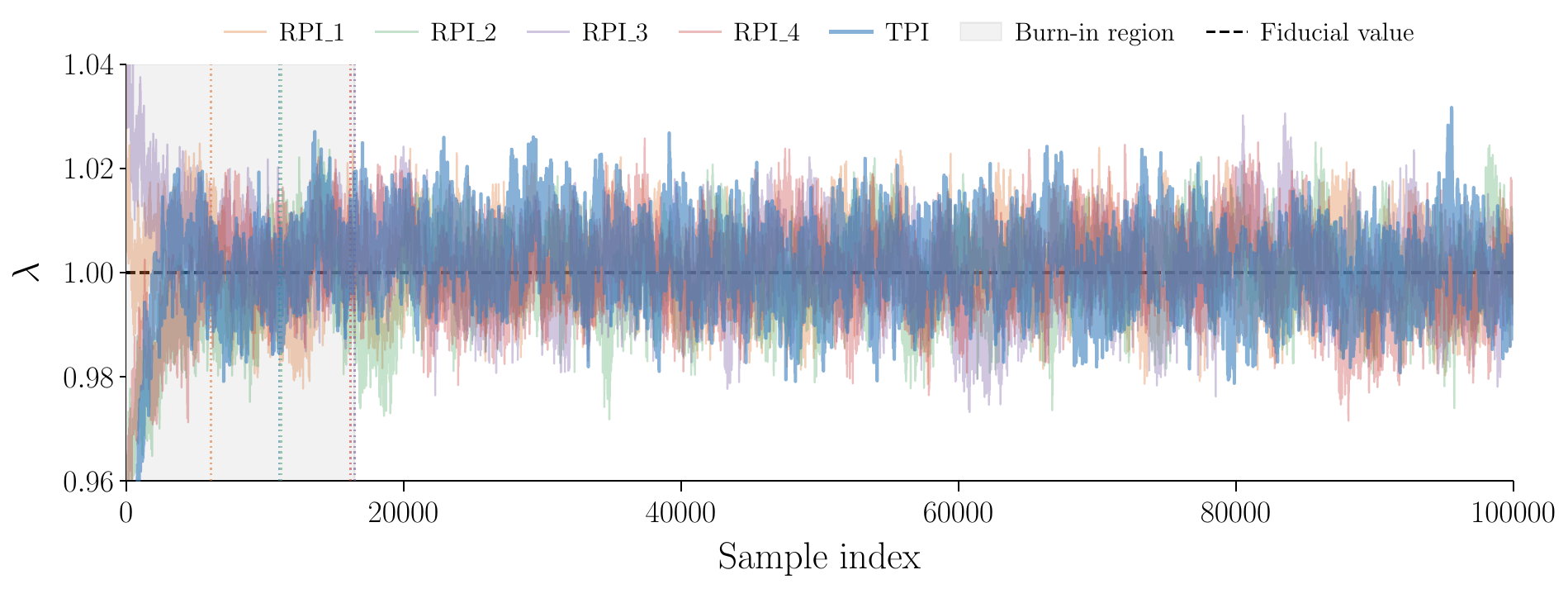}
\caption{\label{fig:trace_plot} Trace plot of $\bdd$ for the mock with $\bdd^\mathrm{fid}=1$. The TPI chain is shown in blue and the four RPI chains in different colors. The dashed black line indicates the fiducial value $\bdd^\mathrm{fid}=1$, and the shaded region marks the burn-in phase that is discarded before the analysis.}
\end{figure}

In this section we describe in detail the analysis of field-level chains. For each mock, we run five Markov chains, one chain initialized at the ground-truth initial conditions (true-phase initialization, TPI), and four chains initialized from independent random initial phases (random-phase initialization, RPI). For each RPI chain, an independent random seed was used to ensure different starting locations in parameter space, as well as different initial values of all the sampled model parameters. Once converged, all chains should sample the same posterior distribution regardless of the initialization. \reffig{trace_plot} shows the trace plot of parameter $\bdd$ for the mock with $\bdd^\mathrm{fid}=1$, with the TPI chain in blue and the four RPI chains in different colors. All chains rapidly converge to a consistent value of $\bdd$ around the fiducial value (dashed black line). The shaded burn-in region is discarded as described below.

\paragraph{Correlation lengths and burn-in.} 
Samples within an MCMC chain are correlated, which reduces the effective number of independent samples. The integrated auto-correlation time quantifies the degree of correlation in the chain and is used to estimate both the number of effective samples, $N_\mathrm{eff}$, and the burn-in length.

For a chain $\{f_s\}_{s=1,\dots,N}$ of some parameter $f$, with sample mean $\langle f_s \rangle = N^{-1} \sum_{s=1}^N f_s$, the auto-correlation function at lag $\Delta t$ is
\begin{equation}
\mathcal{A}(\Delta t) = \frac{1}{N - \Delta t} \sum_{s=1}^{N - \Delta t} f_s\, f_{s+\Delta t} - \langle f_s \rangle^2,
\end{equation}
where $\Delta t \in \{0,1,2,\dots\}$ is the separation between samples. The normalized auto-correlation function is then defined as
\begin{align}
  p(\Delta t)=\frac{\mathcal{A}(\Delta t)}{\mathcal{A}(0)},
\end{align}
so that $p(0)=1$.  The integrated auto-correlation time, or the correlation length of the chain, is estimated from
\begin{align}
  \tau(M)=\sum_{\Delta t= - M}^{M} p(\Delta t) = 1+2 \sum_{\Delta t =1}^M p(\Delta t),
\end{align}
where the window size $M$ is chosen adaptively following Sokal's automatic windowing procedure ~\cite{Sokal_1997} to truncate the sum once it becomes noise-dominated. For each chain, we take $\hat\tau_\mathrm{max}$ to be the largest auto-correlation time across the sampled parameters and discard the first $5\hat\tau_\mathrm{max}$ samples as burn-in. The effective number of post-burn-in samples is $N_\mathrm{post\text{-}burn\text{-}in}/\hat\tau_\mathrm{max}$.

\paragraph{Combined chain.} After burn-in is removed from each chain individually, we concatenate the post-burn-in segments of all five chains into a single combined chain. We extend the runs until this combined chain contains at least 100 effective samples of $\bdd$ per mock. \reftable{tau_eff_samples_all_free} lists the correlation lengths and effective sample sizes of each individual chain, together with the $N_\mathrm{eff}$ of the combined chain used in the main analysis.

\begin{table}[ht]
\centering
\begin{tabular}{l c c c c}
\toprule
Mock & Chain & $\tau_{\bdd}$ & $N_{\mathrm{eff},\bdd}$ & Combined $N_{\mathrm{eff},\bdd}$ \\
\midrule
\multirow{5}{*}{$\bdd^\mathrm{fid} = 0$}
 & TPI    & 12  & 21723 & \multirow{5}{*}{117111} \\
 & RPI\_1 & 22  & 13575 & \\
 & RPI\_2 & 11  & 27196 & \\
 & RPI\_3 & 19  & 16540 & \\
 & RPI\_4 & 331 & 934   & \\
\midrule
\multirow{5}{*}{$\bdd^\mathrm{fid} = 0.25$}
 & TPI    & 2684 & 137  & \multirow{5}{*}{7931} \\
 & RPI\_1 & 137  & 2776 & \\
 & RPI\_2 & 102  & 3680 & \\
 & RPI\_3 & 165  & 2299 & \\
 & RPI\_4 & 116  & 3253 & \\
\midrule
\multirow{5}{*}{$\bdd^\mathrm{fid} = 0.5$}
 & TPI    & 2800 & 67   & \multirow{5}{*}{4067} \\
 & RPI\_1 & 252  & 1315 & \\
 & RPI\_2 & 403  & 769  & \\
 & RPI\_3 & 235  & 1424 & \\
 & RPI\_4 & 903  & 352  & \\
\midrule
\multirow{5}{*}{$\bdd^\mathrm{fid} = 0.75$}
 & TPI    & 912   & 415  & \multirow{5}{*}{2503} \\
 & RPI\_1 & 46072 & 21   & \\
 & RPI\_2 & 42059 & 23   & \\
 & RPI\_3 & 427   & 1036 & \\
 & RPI\_4 & 368   & 1053 & \\
\midrule
\multirow{5}{*}{$\bdd^\mathrm{fid} = 1$}
 & TPI    & 2209 & 156 & \multirow{5}{*}{417} \\
 & RPI\_1 & 1223 & 234 & \\
 & RPI\_2 & 2239 & 225 & \\
 & RPI\_3 & 3297 & 159 & \\
 & RPI\_4 & 3236 & 158 & \\
\midrule
\multirow{5}{*}{$\bdd^\mathrm{fid} = 1.25$}
 & TPI    & 8635  & 89  & \multirow{5}{*}{187} \\
 & RPI\_1 & 19311 & 54  & \\
 & RPI\_2 & 26374 & 42  & \\
 & RPI\_3 & 27628 & 36  & \\
 & RPI\_4 & 9209  & 113 & \\
\bottomrule
\end{tabular}
\caption{Per-chain auto-correlation times $\tau_{\bdd}$ and effective
sample sizes $N_{\mathrm{eff},\bdd}$ of the $\bdd$ parameter, together with the effective sample size of the combined chain used in the main analysis, across mocks with different fiducial values of $\bdd$.}\label{tab:tau_eff_samples_all_free}
\end{table}

The same convergence criterion, i.e. at least 100 effective samples of $\bdd$ in the chain, was applied to the fixed $\bd$ and fixed $\s_\varepsilon$ FLI runs used in \reffig{fixed_params}, and to the lower $\s_\varepsilon$ FLI runs in \refapp{lower_noise}.

\paragraph{Gelman--Rubin diagnostic.}
As an additional convergence check, we compute the Gelman--Rubin statistic $\hat R$~\cite{Gelman_1992} across the five chains for each mock, using the implementation in \texttt{numpyro}. The results are summarized in \reftable{gelman_rubin}. For all mocks and all three parameters, we find $\hat R - 1 \lesssim 0.02$, indicating that the chains have converged to a common distribution. The values are smallest at low $\bdd^\mathrm{fid}$ and grow mildly with $\bdd^\mathrm{fid}$, mirroring the increase in auto-correlation times in \reftable{tau_eff_samples_all_free}.

\begin{table}[ht]
\centering
\begin{tabular}{l c c c}
\toprule
Mock & \multicolumn{3}{c}{Gelman--Rubin $\hat{R}$} \\
\cmidrule(lr){2-4}
 & $\bd$ & $\bdd$ & $\s_\varepsilon$ \\
\midrule
$\bdd^\mathrm{fid} = 0$    & 1.000316 & 1.000013 & 1.000308 \\
$\bdd^\mathrm{fid} = 0.25$ & 1.001398 & 1.000545 & 1.001574 \\
$\bdd^\mathrm{fid} = 0.5$  & 1.000479 & 1.000397 & 1.000576 \\
$\bdd^\mathrm{fid} = 0.75$ & 1.003762 & 1.002913 & 1.003728 \\
$\bdd^\mathrm{fid} = 1$    & 1.008042 & 1.014429 & 1.003166 \\
$\bdd^\mathrm{fid} = 1.25$ & 1.010665 & 1.014537 & 1.000939 \\
\bottomrule
\end{tabular}
\caption{Gelman--Rubin diagnostic $\hat{R}$ computed across the five
chains, for parameters $\bd$, $\bdd$, and $\s_\varepsilon$, across mocks with different fiducial values of $\bdd$.}\label{tab:gelman_rubin}
\end{table}

\subsection{SBI}\label{appendix:SBI_validation}
In this section, we analyze and describe the computational efficiency of composite operator-correlators (OCs) relative to the $n$-point functions, the convergence of the simulation-based inference (SBI) pipeline with simulation budget, and the hyperparameter optimization used to fix the neural posterior estimation (NPE) configuration.

Throughout this appendix, the ``baseline architecture'' refers to the fixed network of \cite{Tucci_2024} introduced in \refsec{methods}:  a Masked Autoregressive Flow (MAF) with 5 autoregressive transforms of 2 hidden layers of 50 units each, trained with learning rate $5\times10^{-4}$ and batch size 50. We use it as a common reference point for the simulation-budget study, and as the starting point for the per-summary optimization described below, which supersedes it.

\paragraph{Data-vector size.}
As discussed in \refsec{OCvsPB}, OCs offer a significant computational advantage over direct estimation of higher-order $n$-point functions, both at the level of the data-vector dimensionality and at the level of the estimator cost.

The galaxy power spectrum is a one-dimensional function of $|\boldsymbol{k}|$ and contributes $N_\mathrm{bin}$ entries to the data vector. Higher $n$-point functions live on much larger grids: the bispectrum on a 3D grid of triangle bins $(k_1, k_2, k_3)$ satisfying the triangle inequality, the trispectrum on a still higher-dimensional grid of tetrahedral configurations, and so on for higher orders. The OC data vector, by contrast, contains at most $n(n+1)/2$ auto- and cross-spectra of operators up to order $n$, each binned in a single wavenumber as for the power spectrum, so its size grows much more slowly with $n$ than for $n$-point functions.

For a cubic box of side length $L = 2000~h^{-1}\mathrm{Mpc}$, the fundamental wavenumber equals $k_f=\frac{2\pi}{L}\approx0.00314~h\,\mathrm{Mpc}^{-1}$. We set $k_\mathrm{max}\equiv\Lambda=0.14 ~h\,\mathrm{Mpc}^{-1}$ and use $N_\mathrm{bin} = 20$ linearly-spaced $k$-bins for the power spectrum, giving an average bin width $\Delta k \approx 2k_f$. The power spectrum data vector then has a size of $N_\mathrm{bin}=20$, while the combined P${+}$B data vector has a total size of $945$ bins, due to different triangle configurations of the bispectrum. The OC data vectors instead have $N_\mathrm{bin}$ elements, with $n=2, 3, 4, 6$ for OC$\_$2nd$\_$LO, OC$\_$2nd$\_$full, OC$\_$3rd$\_$LO and OC$\_$3rd$\_$full, i.e.\ 40 to 120 in our configuration, far smaller than the P${+}$B, and roughly two orders of magnitude smaller than a data vector if we were to also include the trispectrum ($\sim15{,}000$). Moreover, the estimator cost is lower for OCs. For the bispectrum we use the FFT-based estimator of \cite{Scoccimarro_1998}: $N_\mathrm{bin}$ inverse FFTs produce $k$-bin-restricted filtered fields $\delta_{k_i}(\boldsymbol{x})$, and the bispectrum at each triangle bin is the real-space integral $\int d^3x\,\delta_{k_1}\delta_{k_2}\delta_{k_3}$, divided by the number of triangle configurations in that bin. The OCs, by contrast, only require constructing each operator $[\delta_g^\Lambda]^n$ pointwise in real space, one FFT per operator, and one 1D cross-power-spectrum measurement per operator pair. The OC data vectors are correspondingly cheaper to generate, which is an important cost when using SBI, where large simulation suites are required.

\begin{figure}[tbp]
\centering
\includegraphics[width=\textwidth]{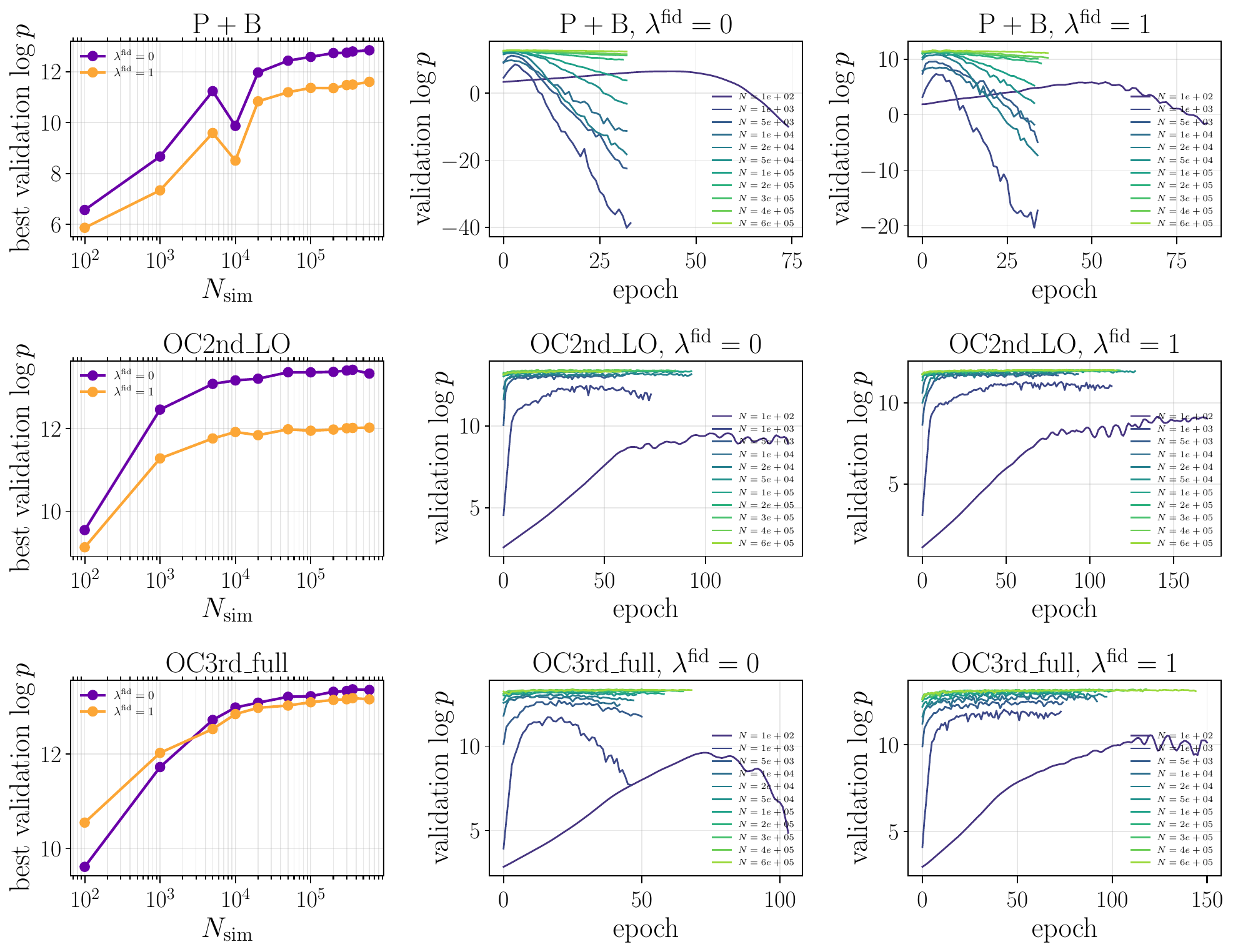}
\caption{\label{fig:nsim_sweep}
Dependence of the trained NPE on the simulation budget, $N_{\mathrm{sim}}$, for P${+}$B, OC$\_$2nd$\_$LO, and OC$\_$3rd$\_$full, using the baseline architecture defined in the text. \emph{Left:} best validation log-probability over all epochs as a function of $N_{\mathrm{sim}}$, for fiducial mocks $\bdd^{\mathrm{fid}} = 0$ (purple) and $\bdd^{\mathrm{fid}} = 1$ (orange). \emph{Center, right:} per-epoch validation log-probability for $\bdd^{\mathrm{fid}} = 0$ and $\bdd^{\mathrm{fid}} = 1$, respectively, color-coded by $N_{\mathrm{sim}}$. The OC summaries plateau at a markedly lower $N_{\mathrm{sim}}$ than P${+}$B, which continues to improve up to $N_{\mathrm{sim}} \sim 10^5$.}
\end{figure}

\paragraph{Convergence with simulation budget.}
\reffig{nsim_sweep} shows how the trained NPE depends on the number of training simulations, $N_{\mathrm{sim}}$, using the baseline architecture defined above. The left column shows the best validation log-probability reached over all training epochs as a function of $N_{\mathrm{sim}}$, and the center and right columns show the per-epoch validation curves at fixed $N_{\mathrm{sim}}$, for the fiducial mocks $\bdd^{\mathrm{fid}} = 0$ and $1$.

The OC summaries reach a stable validation log-probability already at $N_{\mathrm{sim}} \sim 5 \times 10^4$, beyond which additional simulations bring negligible improvement. The P${+}$B summary, by contrast, continues to improve up to $N_{\mathrm{sim}} \sim 10^5$ and, at low budgets, shows pronounced overfitting: the per-epoch validation curves turn over and diverge. This behavior motivates the simulation budgets used in the hyperparameter search below: $N_{\mathrm{sim}} = 10^5$ for P${+}$B and $5 \times 10^4$ for the OCs, so that each summary is optimized in its converged regime while keeping the OC studies, which converge earlier, less costly.

The final production training is carried out at a larger budget, $N_{\mathrm{sim}} = 6 \times 10^5$ for P${+}$B and $N_{\mathrm{sim}} = 3.6\times 10^5$ for OC summaries, comfortably within the plateau.

\paragraph{Per-summary hyperparameter optimization.}
As motivated in \refsec{methods}, we optimize the NPE hyperparameters separately for each of the five summary statistics. We use \textsc{Optuna} \cite{akiba2019optuna} with a tree-structured Parzen estimator sampler to search for a hyperparameter configuration that minimizes the NPE validation loss (the negative log-probability of $(\theta, x)$ pairs under the trained flow) at fixed simulation budget. For each summary the search is run at three fiducial values $\bdd^\mathrm{fid} \in \{0,\,0.5,\,1\}$ to ensure the selected configuration is not tuned to a single value.  The hyperparameters varied are: the learning rate, the number of hidden features, the number of transforms, the number of blocks, training batch size, and the density-estimator. We consider two normalizing flow types for density-estimator: Masked Autoregressive Flows (MAFs) \cite{Papamakarios_2017} with affine transformations, and  Neural Spline Flows (NSFs) \cite{durkan2019neuralsplineflows}, which replace the affine transformations with more flexible monotonic rational-quadratic splines. The models are trained via stochastic optimization of the loss using the Adam optimizer \cite{Kingma_2014}. Search ranges are given in \reftable{optuna_search}, dropout is disabled and the gradient-norm clipping is fixed at $6.0$. Each NPE training trial halts after $30$ epochs without validation improvement and reports its final validation loss to the sampler. We run 50 trials per (summary, $\bdd^\mathrm{fid}$) cell.

\begin{table}[ht]
\centering
\begin{tabular}{l l}
\toprule
Hyperparameter & Search range \\
\midrule
Learning rate                & $[10^{-6},\,10^{-2}]$, log-uniform \\
Hidden features              & $[10,\,100]$, integer \\
Number of transforms         & $[3,\,13]$, integer \\
Number of blocks             & $[1,\,7]$, integer \\
Training batch size          & $\{128,\,256,\,512,\,1024\}$ \\
Density estimator            & MAF or NSF \\
\midrule
Dropout probability          & $0$ (fixed) \\
Batch normalization          & off (fixed) \\
Gradient clipping (max norm) & $6.0$ (fixed) \\
Validation fraction          & $0.1$ (fixed) \\
\bottomrule
\end{tabular}
\caption{Hyperparameters varied by \textsc{Optuna} and their search ranges. Each of the 15 studies (5 summaries $\times$ 3 values of $\bdd^\mathrm{fid}$) ran 50 trials. The lower block lists settings held fixed across all trials.}\label{tab:optuna_search}
\end{table}

The validation loss of a trained NPE is not deterministic in the hyperparameters: random weight initialization and stochastic mini-batch ordering induce a run-to-run scatter even with hyperparameters, training data, and prior all held fixed. We quantify this ``seed noise'' $\s_\mathrm{seed}$ by retraining a fixed configuration with 10 distinct random seeds and taking the standard deviation of the resulting validation losses. A gap in validation loss smaller than $\s_\mathrm{seed}$ between two Optuna trials is uninformative: the apparently better trial is better only by chance.

This sets a floor on the useful simulation budget. At a pilot budget of $N_\mathrm{sim}=2\times10^4$ we find $\s_\mathrm{seed}$ comparable to, or larger than, the spread among the best-scoring trials in a majority of cells, so the search at that budget cannot reliably rank architectures. Raising the budget both reduces $\s_\mathrm{seed}$ and widens the gap between good and poor configurations. We therefore set the search budget per summary so that the trial-to-trial spread exceeds $\s_\mathrm{seed}$: $N_\mathrm{sim}=5\times10^4$ for the OC summaries, already in the plateau identified in the convergence study above, and $N_\mathrm{sim}=10^5$ for P${+}$B, for which the validation loss continues to improve up to that budget. 

To convert \textsc{Optuna} trials into a single production configuration per summary, we pool  the trials across the three $\bdd^\mathrm{fid}\in\{0,\,0.5,\,1\}$ values, take the top 10 by validation loss, and report the median of each continuous  hyperparameter and the mode of the categorical density-estimator choice. Pooling across $\bdd^\mathrm{fid}$ avoids tuning the network to a single fiducial point, while taking the top-10 (rather than the single best) suppresses sensitivity to the trial-to-trial seed scatter $\s_\mathrm{seed}$ identified above. The selected per-summary configurations are listed in \reftable{optuna_picks}. The density estimator selected is NSF for all five summaries; the optimal learning rate spans $6.3\times10^{-5}$ to $3.5\times10^{-4}$ for the OC summaries and is $7.5\times10^{-6}$ for P${+}$B; the number of hidden features ranges from 13 to 91. These configurations are used for production NPE training in \refsec{results}, where the final production training uses $N_\mathrm{sim}=3.6\times10^5$ for OCs and $N_\mathrm{sim}=6\times10^5$ for P${+}$B, well within the converged regime.

\begin{table}[ht]
\centering
\begin{tabular}{l c c c c c c}
\toprule
Summary & Learning rate & Hidden feat. & Transforms & Blocks & Batch size & Estimator \\
\midrule
P${+}$B               & $7.49\times10^{-6}$ & 91 & 5 & 1 & 128 & NSF \\
OC$\_$2nd$\_$LO   & $2.03\times10^{-4}$ & 35 & 5 & 4 & 256 & NSF \\
OC$\_$2nd$\_$full & $3.51\times10^{-4}$ & 14 & 6 & 5 & 512 & NSF \\
OC$\_$3rd$\_$LO   & $1.23\times10^{-4}$ & 13 & 5 & 4 & 256 & NSF \\
OC$\_$3rd$\_$full & $6.30\times10^{-5}$ & 52 & 6 & 1 & 256 & NSF \\
\bottomrule
\end{tabular}
\caption{Production NPE configuration per summary, obtained as the median (continuous) or mode (categorical) over the top-10 \textsc{Optuna} trials (ranked by validation loss) pooled across $\bdd^\mathrm{fid}\in\{0,\,0.5,\,1\}$.}\label{tab:optuna_picks}
\end{table}

\paragraph{Production widths and their seed uncertainty.}
For the production posteriors we quantify both the parameter constraint and its sensitivity to training stochasticity by retraining each $(\text{summary},\,\bdd^\mathrm{fid})$-configuration with 10 independent random seeds (random weight initialization and mini-batch ordering), holding the training data, prior, and observed data vector fixed. For each seed we draw the posterior and measure the standard deviation of each parameter's marginal posterior, i.e. its width. We then report two complementary summaries: the seed-averaged width, obtained by averaging the per-seed widths, with an uncertainty given by their scatter across seeds; and the pooled width, obtained by concatenating the samples of all seeds into a single set and taking its standard deviation. The seed-averaged width is the quantity plotted throughout \refsec{results}, with the seed scatter shown as its error bar; the pooled width provides a cross-check that additionally folds in the seed-to-seed shifts of the posterior mean. The pooled posterior samples themselves are what we display in the corner plots of \reffig{posteriors_PB}. This concatenation is of posterior samples, and is distinct from the concatenation of training simulations described in \refsec{methods}.

\begin{figure}[tbp]
\centering
\includegraphics[width=\textwidth]{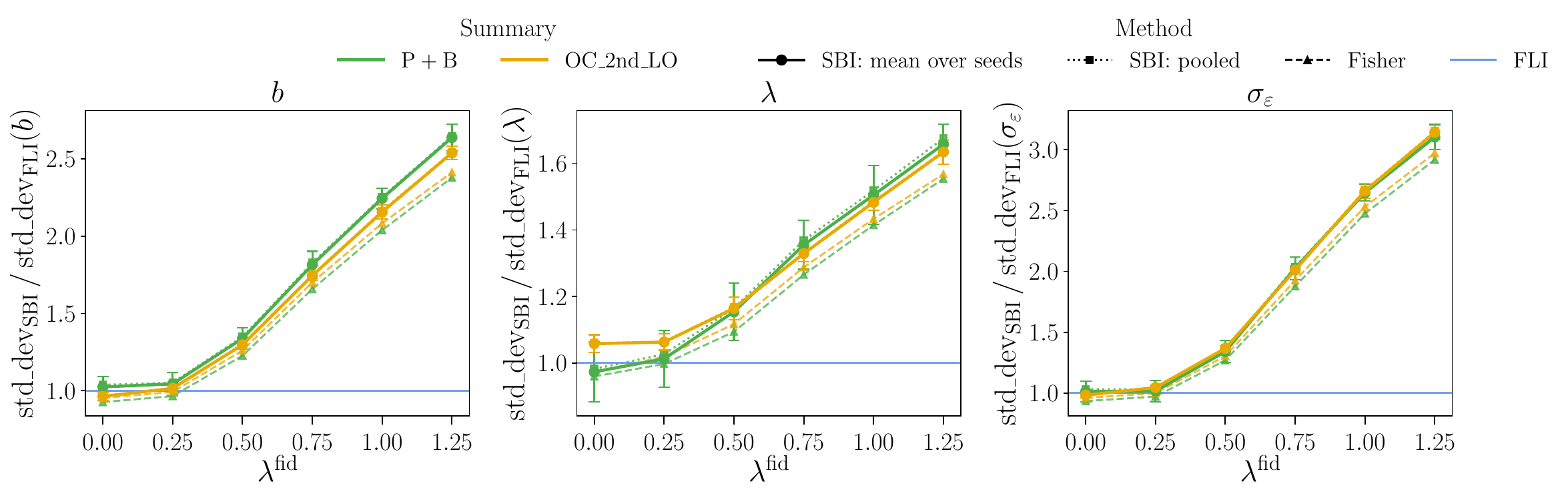}
\caption{\label{fig:seed_validation}
Per-parameter posterior widths for P${+}$B (green) and OC$\_$2nd$\_$LO (yellow), normalized by the field-level result, as a function of $\bdd^\mathrm{fid}$, for the three
sampled parameters: $\bd$ \emph{(left)}, $\bdd$ \emph{(center)}, and
$\s_\varepsilon$ \emph{(right)}. Solid lines with circles show the seed-averaged width with
error bars given by the scatter across seeds; dotted lines with squares show the pooled
width; dashed lines with triangles show the Fisher prediction. The horizontal line at unity
marks the field-level reference. The seed-averaged and pooled estimates agree closely,
indicating that seed-to-seed shifts of the posterior mean are subdominant to the width
itself.}
\end{figure}

\reffig{seed_validation} shows the normalized widths for all three parameters for P${+}$B (green) and OC$\_$2nd$\_$LO (yellow). At $\bdd^\mathrm{fid}=0$ the summary and field-level widths agree for every parameter (ratios consistent with unity within the seed scatter), as expected in the Gaussian limit where P${+}$B and OC$\_$2nd$\_$LO are equivalent to the field-level constraints. As $\bdd^\mathrm{fid}$ increases, the summary widths grow relative to FLI for all parameters, with the steepest growth in $\s_\varepsilon$ and the mildest in $\bdd$. Across the whole range the seed-averaged widths track the Fisher predictions (dashed) for both summaries, lying marginally above them by an amount comparable to the seed scatter. This confirms that the production NPE recovers the Fisher information to within the precision set by training stochasticity, for every parameter.

\section{Subtracting the disconnected part}\label{appendix:disconnected}
The auto-correlator $\langle (\d_g^2) (\d_g^2)\rangle$ contains a Gaussian disconnected part that is fully determined by the two-point function $\langle \d_g \d_g \rangle$ and therefore carries no information beyond it. To see this, consider the mean-subtracted quadratic operator $O^{(2)}(\boldsymbol{x}) = \d_g^2(\boldsymbol{x}) - \langle\d_g^2\rangle$ (the mean subtraction is applied in our measurement). Treating $\d_g$ as Gaussian, Wick's theorem gives
\begin{equation}
\langle \d_g^2(\boldsymbol{x})\,\d_g^2(\boldsymbol{y})\rangle = \langle\d_g^2\rangle^2 + 2\,\xi(\boldsymbol{x}-\boldsymbol{y})^2\,,
\label{eq:wick_disc}
\end{equation}
where $\xi(r) =\langle\d_g(\boldsymbol{x})\d_g(\boldsymbol{x}+\boldsymbol{r})\rangle$ is the two-point correlation function. Thus, in the Gaussian limit, $\langle O^{(2)}(\boldsymbol{x})\,O^{(2)}(\boldsymbol{y})\rangle = 2\,\xi(r)^2$, and the Gaussian disconnected contribution to the spectrum $\langle(\d_g^2)(\vk)(\d_g^2)(-\vk)\rangle$ is its Fourier transform, $\mathrm{FT}[\,2\,\xi(r)^2\,](k)$.
While the presence of this term is in principle not an issue for SBI, we expect better performance if the elements of the data vector are less correlated. Hence, we calculate and subtract this Gaussian disconnected part in our analysis. Specifically, from the sharp-$k$-filtered galaxy field $\d_g$ (the same field entering the measured correlator) we measure its auto-power spectrum $P_{\d_g}(k)$, build a log--log spline of $P_{\d_g}(k)$ in $k$, and use it to populate a 3D Fourier-space grid out to the cutoff $\Lambda$. We transform this grid to real space to obtain $\xi(r)$, form $2\,\xi(r)^2$, and transform back to Fourier space. The result is binned in the same $k$-bins as the data vector, ensuring consistent bin averaging, and subtracted from the measured $\langle \d_g^2 \d_g^2\rangle$ bin by bin. We use $N_\mathrm{disc} = 20$ linearly spaced bins over $[3 k_f, \Lambda]$ for the intermediate $P_{\d_g}(k)$ spline. The result is mildly sensitive to this choice due to sample fluctuations, so we deliberately keep $N_\mathrm{disc}$ small.

An analogous decorrelation is applied at cubic order. We do not subtract a disconnected part from a measured spectrum in this case; instead we Wick-order the cubic operator itself, replacing $\d_g^3 \to \d_g^3 - 3\langle\d_g^2\rangle\,\d_g$. For Gaussian $\d_g$ this combination satisfies $\langle\d_g(\boldsymbol{x})\,[\d_g^3 - 3\langle\d_g^2\rangle\d_g](\boldsymbol{y})\rangle = 0$, i.e. it removes the piece of $\d_g^3$ that is linearly correlated with $\d_g$ (and hence redundant with the power spectrum).

\begin{figure}[tbp]
\centering 
\includegraphics[width=0.9\textwidth]{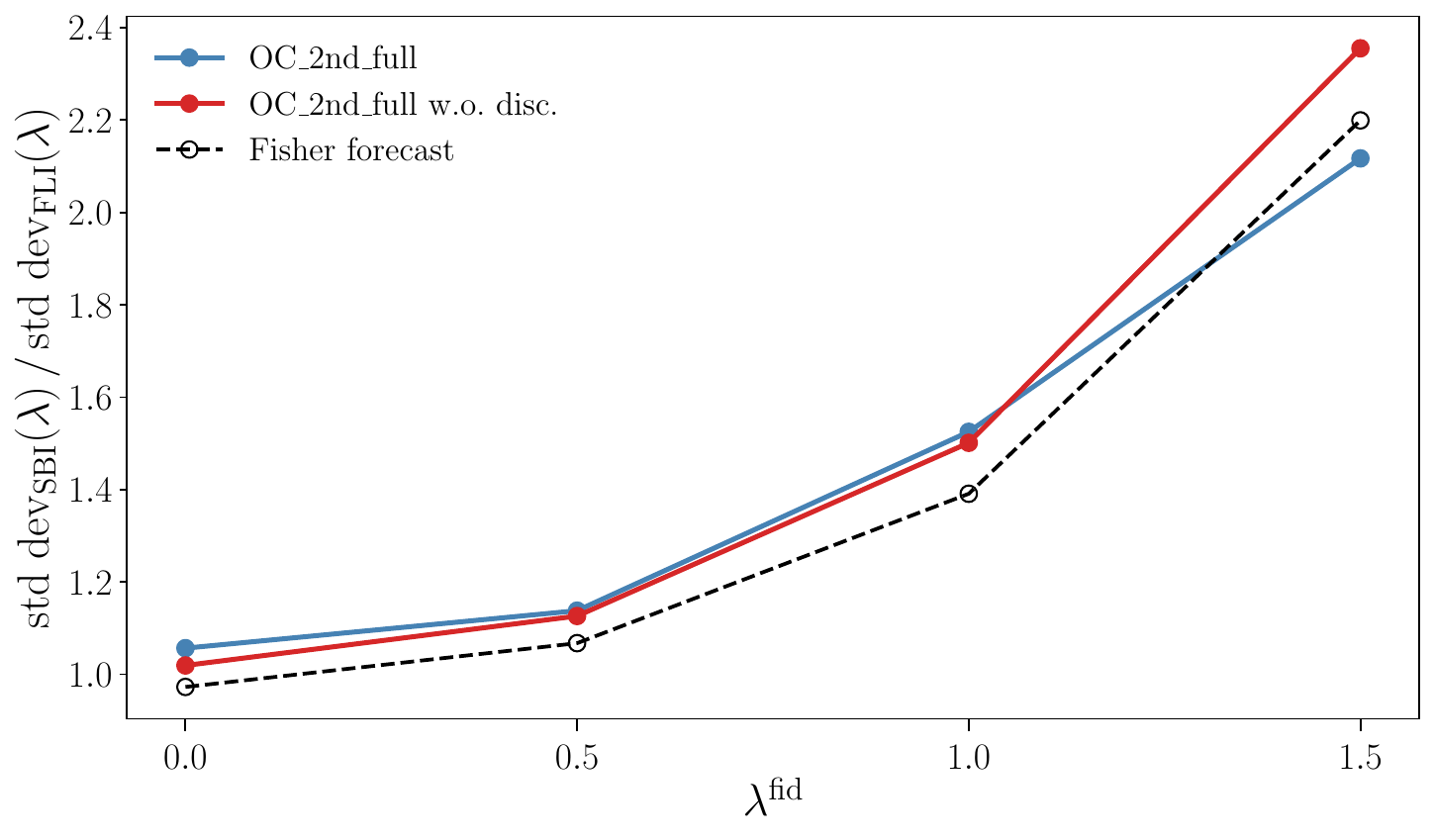}
\caption{\label{fig:disconnected} Inferred 68\% CL error bar on $\bdd$ for OC$\_$2nd$\_$full as a function of $\bdd^\mathrm{fid}$, with all parameters sampled, comparing the full correlator (``full'') with the version where the Gaussian disconnected part of $\langle\d_g^2\d_g^2\rangle$ is subtracted (``w.o. disc.'').}
\end{figure}

\reffig{disconnected} compares the inferred 68\% CL error bar on $\bdd$ from OC$\_$2nd$\_$full with and without subtracting the disconnected part. The curves labeled ``w.o. disc.'' have the disconnected part removed. The curves shown are each from a single NPE seed; the seed-to-seed scatter on each is comparable to the gap between them, so the choice of subtraction has no measurable impact on our main conclusions. We use the subtracted version (``w.o. disc.'') throughout.

\section{Lower noise}\label{appendix:lower_noise}
\begin{figure}[tbp]
\centering 
\includegraphics[width=\textwidth]{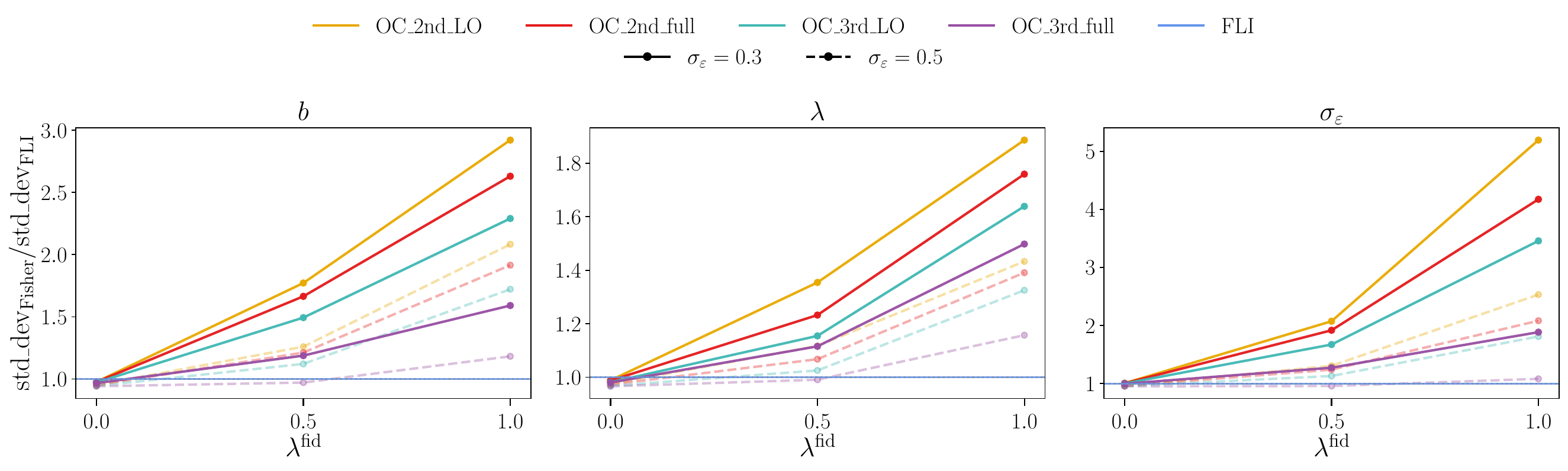}
\caption{\label{fig:lower_sigma}
Fisher forecast for the 68\% CL uncertainty on each parameter, normalized by the field-level result, as a function of $\bdd^\mathrm{fid}$, for a lower stochastic amplitude $\s_\varepsilon = 0.3$ (solid lines) compared to the fiducial $\s_\varepsilon = 0.5$ (dashed lines). Colors correspond to OC combinations as labeled. At lower noise, the gap between summary statistics and field-level inference grows more rapidly with $\bdd^\mathrm{fid}$, reflecting  the increased importance of higher-order information when the noise floor is lower. All parameters are inferred jointly.}
\end{figure}
\reffig{lower_sigma} shows the Fisher-predicted 68\% CL uncertainty ratios for $\s_\varepsilon=0.3$ (solid) alongside the fiducial $\s_\varepsilon=0.5$ (dashed), for all four OC summaries.  At lower noise, the signal-to-noise per mode is higher, so the data carry more information, but this also means that the information lost by using a compressed summary rather than the full field is amplified. Consequently, the ratios grow more steeply with $\bdd^\mathrm{fid}$ at $\s_\varepsilon=0.3$ than at $\s_\varepsilon=0.5$ for all parameters and all summaries. Concretely, at $\bdd^\mathrm{fid}=1$ the OC$\_$2nd$\_$LO uncertainty on $\bdd$ rises from $\sim 45\%$ above FLI at $\s_\varepsilon=0.5$ to $\sim 90\%$ above FLI at $\s_\varepsilon=0.3$, while OC$\_$3rd$\_$full rises from $\sim 15\%$ to $\sim 50\%$ above FLI over the same noise change; the same trend, with a comparable relative increase, is visible for $\bd$, and is most pronounced for $\s_\varepsilon$ itself. The qualitative hierarchy between summaries observed at $\s_\varepsilon=0.5$ is preserved at $\s_\varepsilon=0.3$: OC$\_$2nd$\_$full and OC$\_$3rd$\_$LO sit between OC$\_$2nd$\_$LO and OC$\_$3rd$\_$full, with OC$\_$3rd$\_$full remaining closest to the field-level result. At $\bdd^\mathrm{fid}=0$ the information on $\bdd$ is fully captured by P${+}$B (or equivalently OC$\_$2nd$\_$LO); higher-order OCs add no further information in this Gaussian limit, and the Fisher ratios are consistent with unity. Overall, the gap in information gain between field-level inference and summary statistics widens at lower noise, motivating the use of field-level or higher-order OC methods for future, high signal-to-noise datasets.

\bibliographystyle{JHEP}
\bibliography{bibliography}

\end{document}